\documentclass[12pt]{article}
\usepackage[margin=2.5cm]{geometry}  % add also "show frame" options  
\usepackage{graphicx,psfrag,epsf}
\usepackage{enumerate}
\usepackage{url} % not crucial - just used below for the URL 
\usepackage{float}
\usepackage{xcolor}
\usepackage{algorithm}
\usepackage{algorithmic}
\usepackage{amsmath}
\usepackage{amssymb}
\usepackage{graphicx}
\usepackage{subcaption}
\usepackage{natbib}
\usepackage{lineno}
\usepackage{dcolumn}
\usepackage{multirow}
\usepackage{comment}
\usepackage{units}
\usepackage{booktabs}
\usepackage{colortbl}
\usepackage{rotating}
\usepackage{epstopdf}
\usepackage[table]{xcolor}  
\newcolumntype{.}{D{.}{.}{-1}}
\newcolumntype{d}[1]{D{.}{.}{#1}}
\usepackage{hyperref}
\hypersetup{
	colorlinks=true,
	citecolor=blue,
}

\usepackage{rotating}
\usepackage{epsfig}%
\usepackage{epstopdf}
\def \id{1\hskip -3pt \mbox{l}}

\def\btheta{\mbox{\boldmath $\theta$}}

\def\bf{\mbox{\boldmath $f$}}

\def\1{\mbox{1}}

\def\btheta{\mbox{\boldmath $\theta$}}

\def\btheta{\mbox{\boldmath $\theta$}}

\def\0{\mbox{\bf{0}}}

\def\0{\mbox{\bf{0}}}

\def\0{\mbox{\bf{0}}}

\newcommand{\nin}{\noindent}

\def\bkR{{\rm I\kern-.17em R}}

\def \id{1\hskip -3pt \mbox{l}}
\def \1n{1\hskip -3pt \mbox{N}}

\def \Frac {\displaystyle \frac }

\newfont{\bbf}{cmbx12 scaled 1435}
\usepackage[normalem]{ulem}

\begin{document}

\setlength{\baselineskip}{.26in}
\thispagestyle{empty}
\renewcommand{\thefootnote}{\fnsymbol{footnote}}
\vspace*{0cm}
\begin{center}

\setlength{\baselineskip}{.32in}
{\bbf Bubble Detection with Application to Green Bubbles: A Noncausal Approach
}\\

\vspace{0.5in}

\large{Francesco Giancaterini}\footnote{Centre for Economic and International Studies, Tor Vergata University of Rome, Italy, email: {\it francesco.giancaterini@uniroma2.it}}\footnote{Ugo Bordoni Foundation, Rome, Italy, email: {\it fgiancaterini@fub.it}}, 
\large{Alain Hecq}\footnote{Department of Quantitative Economics, Maastricht University, The Netherlands, email: {\it a.hecq@maastrichtuniversity.nl}}, 
\large{Joann Jasiak}\footnote{Department of Economics, York University, Canada, e-mail: {\it jasiakj@yorku.ca}}, 
\large{Aryan Manafi Neyazi }\footnote{Department of Economics, York University, Canada, e-mail: {\it aryanmn@yorku.ca} \\ 
The authors acknowledge the financial support of the Natural Sciences and Engineering Research Council (NSERC) and the Mathematics of Information Technology and Complex Systems (MITACS) of Canada, as well as the financial support of the Italian Ministry of University and Research (MUR) under the PRIN 2022 grant no. 20223725WE. The authors are grateful to C. Gourieroux and P. Sadorsky for their valuable comments. The paper was presented at the Intermediate Workshop of the PRIN 2022 project, the CFE–CMStatistics Conference (2024), the 1st Workshop on Non-causal Econometrics (2025), and the annual meetings of the Statistical Society of Canada (SSC), the Canadian Econometric Study Group (CESG) in 2025 and the XVI Workshop in Time Series Econometrics in 2026 in Zaragoza.}

\setlength{\baselineskip}{.26in}
\vspace{0.4in}

\today\\

\medskip

\vspace{0.3in}
\begin{minipage}[t]{12cm}
\small
\begin{center}
Abstract \\
\end{center}
This paper introduces a new approach for bubble detection based on mixed causal and noncausal autoregressive processes and their tail process representation during an explosive episode. Departing from traditional definitions of bubbles as nonstationary and temporarily explosive processes, we adopt a perspective in which prices are assumed to follow a strictly stationary process, with the bubble considered an intrinsic component of its nonlinear dynamics. The proposed approach provides a bubble indicator for detecting bubbles and measuring their duration. We %illustrate our approach with
implement our strategy to investigate the phenomenon called the "green bubble" in the field of renewable energy investment.

\bigskip


\bigskip


\end{minipage}

\end{center}
\renewcommand{\thefootnote}{\arabic{footnote}}
\newpage
\section{Introduction}
This paper introduces a novel method of bubble detection based on strictly stationary noncausal autoregressive processes and their tail process representation during a bubble. In this context, a simple statistic with an asymptotic normal distribution is proposed to provide an ex ante warning of a potentially emerging bubble. Ex post, this statistic is a tool for measuring the duration of the bubble and identifying its starting and ending dates. The proposed diagnostic method detects single and multiple bubbles with rates of explosion, determined by the noncausal autoregressive coefficients of the process. The bubbles can burst vertically to zero, or decline at a slower rate, depending on the presence of a causal component in the strictly stationary mixed causal-noncausal process. Our approach is also applicable to jump detection in the conventional causal autoregressive processes with heavy-tailed non-Gaussian error distributions. In addition, the proposed method yields simple prediction formulas for the time to peak of a bubble and its duration.

During the bubble growth and/or decline phases, the strictly stationary noncausal autoregressive process admits a tail process representation. The tail process was first defined by \cite{basrak2009regularly} as the weak limit of finite-dimensional distributions of time series, conditional on the occurrence of an extreme value. This definition was further generalized to the spectral tail process by \cite{kulik2020heavy}, which is based solely on normalized observations exceeding a high threshold, and interpreted as "a model for the clusters of exceedances". The tail process representation is valid for geometrically ergodic Markov processes, and consequently also for mixed causal-noncausal autoregressive MAR(1,1) and purely noncausal MAR(0,1) processes, where the notation MAR($r,s$) indicates a mixed model of causal order $r$ and noncausal order $s$ with locally explosive patterns including bubbles and spikes. As shown, for example, by \cite{gourieroux2017local}, \cite{fries2019mixed} and \cite{cavaliere2020bootstrapping}, the MAR(1,1) and MAR(0,1) processes are well-suited for modeling variables such as commodity and cryptocurrency prices. We show that these processes can also represent the time series of green stock prices
that display bubbles and spikes, allowing us to examine bubbles that occurred after year 2020.

The MAR processes have recently received attention in applications to bubble forecasting and testing. \cite{de2025forecasting} consider the forecasts of extreme trajectories of $\alpha$-stable MAR processes, based on a measure describing the conditional distribution of a normalized path of the process, following a large value. The method is applied to predict the occurrences of the El Niño and La Niña phases of the temperatures and winds of the Pacific Ocean. In comparison, our approach is applicable to MAR models with any heavy-tailed error distribution and focuses on detecting rather than predicting the bubbles. In addition, it is computationally simple as it does not require determining the length of a future path as a tuning parameter. 
\cite{Blasq} also consider the $\alpha$-stable distributed processes. They introduce a test for bubbles based on the detection of a large future shock in the forward-looking component of a MAR process and apply it to oil prices. Compared to their approach, our proposed test statistic has the advantage of being easy to compute and having a known asymptotic normal  distribution.

Alternative methods for detecting bubbles, based on the Dickey-Fuller augmented test (e.g., SADF and GSADF), assess unit roots and explosive regimes in nonstationary autoregressive processes  [\cite{phillips2011explosive}, \cite{phillips2015testing}]. Our approach departs from these conventional methods by employing mixed autoregressive causal-noncausal models, which capture locally explosive patterns observed in the data. Unlike those detection approaches, we assume that the series follows a strictly stationary non-Gaussian process in which bubbles are an inherent part of the dynamics, rather than distinguishing a stationary and non-stationary (unit root) regime of a time series. In addition, our approach accommodates local explosive patterns with various explosion and burst rates, which are estimable. 

The rest of the paper is organized as follows. Section \ref{sec:Noncausal} reviews the causal-noncausal autoregressive processes and their estimation methods, with a focus on the semi-parametric Generalized Covariance (GCov) estimator. Section 3 studies the behavior of the MAR process and its tail dynamics during a bubble period. Section 4 develops test statistics to detect the bubbles and determine their duration. In Section 5, we apply our approach to investigate the presence and duration of "green bubbles" in the green energy stock market. Section \ref{sec:Conclusions} concludes.
%We focus on three key financial instruments: the Renewable Energy Industrial Index (Renixx), the WilderHill Clean Energy ETF (WHETF), and the iShares Global Clean Energy ETF (iShares). 
Appendices A and B contain additional technical results, and Appendix C presents the simulation tables discussed in Section 4.3 of the main paper. The following notation is used: $\{y_t, t \in \mathbb{Z}\}$, denotes the strictly stationary mixed autoregressive (MAR) process representing the green stock prices, $y$ denotes a high threshold selected among the admissible values of $y_t$ which can be exceeded at an exogenous date, and $N$ is a random variable representing distance in time to the peak of a bubble.

\section{Causal-Noncausal Processes}
\label{sec:Noncausal}

The causal-noncausal models represent stationary processes characterized by locally explosive patterns, such as bubbles and spikes. The univariate causal-noncausal models were examined, for example, by \cite{breid1991maximum} and \cite{lanne2011noncausal}, and extended to multivariate analysis by \cite{lanne2013noncausal}, \cite{gourieroux2017noncausal, gourieroux2023generalized}, and \cite{davis2020noncausal}. In applied research, causal-noncausal models were used to study various economic and financial variables, including Bitcoin prices [\cite{hencic2015noncausal}, \cite{cavaliere2020bootstrapping}], stock market indices [\cite{gourieroux2017local}], commodity prices [\cite{hecq2021forecasting}, \cite{lof2017noncausality}], and inflation rates [\cite{lanne2013noncausal}, \cite{hecq2023predicting}].
The main advantage of these models is their ability to capture complex nonlinear patterns, such as local trends and conditional heteroskedasticity, while still resembling traditional linear time series models in terms of the specification. However, the standard Box–Jenkins approach to identifying and estimating linear time series processes does not apply here, as it relies on the assumption of Gaussian errors. Under the assumption of Gaussian errors, the causal and noncausal dynamics cannot be distinguished (see \cite{gourieroux2015pricing}). Therefore, for identification, non-Gaussian error distributions are required in the autoregressive causal–noncausal processes.

\subsection{Univariate Causal-Noncausal Models}
\label{sec:Uni}
A strictly stationary univariate mixed causal-noncausal autoregressive MAR($r,s$) model is defined as:
\begin{equation} \label{eq:MAR} 
    \Phi(L)\Psi(L^{-1})y_t = \epsilon_t,
\end{equation}
where the error term $\epsilon_t$ is non-Gaussian, independent, identically distributed (i.i.d.) and such that  $E(|\epsilon_t|^{\delta}) < \infty$ for $\delta > 0$ [Gourieroux and Zakoian (2015)]\footnote{This condition allows the errors to have infinite variance, and possibly infinite mean: for $\delta \geq 2$ the second-order moments exist, for $\delta \in [1,2)$ the errors have infinite variance, but finite first-order moment, for $\delta \in (0,1)$ the errors have no first-order moments. \cite{lanne2011noncausal} assumes errors with zero mean and a finite variance. }. The polynomial $\Phi(L)$ in the lag operator $L$ is of order $r$. The polynomial $\Psi(L^{-1})$ in the leading operator $L^{-1}$ is of order $s$. Both polynomials $\Phi(L)$ and $\Psi(L^{-1})$ have roots outside the unit circle. 
%REMOVED FOR REF 2: such that $E(|\epsilon_t|^{\delta}) < \infty$ for $\delta > 0$ 

The MAR($r,s$) process \eqref{eq:MAR} admits a unique strictly stationary solution, which is a two-sided moving average of order infinity MA($\infty$):
\begin{equation*}
y_t = \sum_{h=-\infty}^{\infty} c_h \epsilon_{t-h},
\end{equation*}
in past, present and future shocks, with $c_0 = 1$  [\cite{breid1991maximum}]. This MA($\infty$) representation exists and is unique, and the coefficients $c_h$ on past and future errors are distinct and uniquely defined, provided that $(\epsilon_t)$ are non-Gaussian [See, \cite{breid1991maximum} and \cite{gourieroux2015uniqueness} for errors with finite variance, and with infinite moments, respectively]. When $y_t$ is purely noncausal (resp. causal), the coefficients $c_h$ are zero for all $h > 0$ (resp. $h < 0$). Thus, a purely causal process is determined only by the past and present shocks, while a purely noncausal process is influenced only by the present and future shocks. For $r=s=1$, we obtain the MAR($1,1$) process:
\begin{equation}
(1- \phi L)(1-\psi L^{-1}) y_t = \epsilon_t,
\end{equation}

\noindent with $|\psi|<1, |\phi|<1$, which is purely causal (resp. noncausal) if $\psi= 0$ (resp. $\phi= 0$). For each of these pure processes, the effects of a large $\epsilon_t$ are easily distinguished, as a large error leads to a (vertical) jump if $\psi = 0$ and $\phi>0$, and an explosive bubble with a (vertical) burst if $\psi > 0$  and $\phi=0$.

The MAR($1,1$) process can be decomposed into the following unobserved  components [\cite{lanne2011noncausal}]:
\begin{equation}
  u_t  := (1- \phi L)  y_t \;\; \mbox{or}, \;\; (1-\psi L^{-1})  u_t = \epsilon_t, 
  \end{equation}
\noindent   and
\begin{equation}
  v_t :=  (1-\psi L^{-1})  y_t \;\; \mbox{or}, \;\; (1- \phi L)   v_t = \epsilon_t.
\end{equation}

\noindent \cite{gourieroux2016filtering} show that $u_t$ is $\epsilon$-noncausal (dependent on the future and present values of $\epsilon$) and $y$-causal (dependent on the past and present values of $y$), representing the regular dynamics of $y_t$. In contrast, $v_t$ is $\epsilon$-causal (dependent on the past and present values of $\epsilon$) and $y$-noncausal (dependent on the future and present values of $y$), representing the explosive part of the process, including bubbles and spikes. Henceforth, $u_t, v_t$ are called the unobserved causal and noncausal components of $y_t$, respectively.

The process $y_t$ has the following deterministic representation based on the above unobserved components that can be used for simulations and bootstrapping:
$$
y_t = \Frac{1}{1-\phi \psi} (\phi v_{t-1} + u_t), \; \mbox{or} \;\;y_t = \Frac{1}{1-\phi \psi} (v_t + \psi u_{t+1}). 
$$

\noindent We observe that $y_t$ is a linear function of the first lag of $v_t$ and of the current value of  $u_t$. Alternatively, $y_t$ can be expressed as a linear function of the current value of $v_t$ and of the first lag of $u_t$.

The autocovariances of the latent components defined in eq. (3) and (4), respectively, help us distinguish the bubble episode in the process $\{y_t\}$. We observe that the autocovariances at lag $h \geq 1$ of $u_{t+h}$ and  $v_t$  conditional on $y_t=y$ are time-varying, in general, provided that $y$ is not an extreme [see Appendix B]. When $y$ is large, their behavior is different and can be examined using the tail process described in Section 3. In practice, the latent components are computed given the estimated values of 
the autoregressive parameters. The estimation methods are discussed in the next section.

\subsection{The GCov Estimator}
\label{sec:GCov}

\indent One way to estimate and identify MAR($r,s$) models is by using a parametric non-Gaussian Maximum Likelihood (ML) approach [see, e.g. \cite{hecq2016identification}]. Alternatively, the semi-parametric Generalized Covariance (GCov) estimator can be used, which does not require any distributional assumptions on the errors other than satisfying the i.i.d. and non-Gaussianity conditions, the latter one being required for identification. The GCov is a one-step estimator that is consistent, asymptotically normally distributed, and semi-parametrically efficient. It can achieve parametric efficiency in special cases [\cite{gourieroux2023generalized}]. The GCov minimizes a portmanteau-type objective function involving nonlinear autocovariances, i.e., the autocovariances of nonlinear transformations of model errors, which successfully identify the causal and noncausal dynamics [\cite{chan2006note}].

Let us consider the nonlinear transformations $a\left(\epsilon_t\right)=a_1(\epsilon_t),...,a_K(\epsilon_t)$ of the error process that increase its dimension from 1 to $K$. These transformations satisfying the regularity conditions given in \cite{gourieroux2023generalized} are introduced to identify the noncausal and nonlinear dynamics and to ensure the existence of moments of the transformed errors.
Let $\hat{\Gamma}^a\left(h; \theta\right), h=1,...,H$ denote the autocovariance matrices of the transformed errors at lags $h=0,...,H$, with $\hat{\Gamma}^a\left(0; \theta\right)$ representing their variance, and $\theta$ the vector of autoregressive parameters. The
GCov estimator $\hat{\theta}_{T}$ minimizes the following objective function:
\begin{equation}
    \hat{\theta}_{T}= \underset{\theta}{\mathrm{argmin}} \sum_{h=1}^{H} Tr \left[ \hat{\Gamma}^a\left(h; \theta\right) \hat{\Gamma}^a\left(0; \theta\right)^{-1}\hat{\Gamma}^a\left(h; \theta\right)' \hat{\Gamma}^a\left(0; \theta\right)^{-1} \right],
    \label{eq:GCov22}
\end{equation}
\nin where $Tr$ denotes the trace of a matrix and $L_T( \hat{\theta}_T,H)$ is the value of the objective function at its minimum\footnote{When the number of transformations $K$ is large and the inversion of the variance matrix becomes difficult, we can replace $\hat{\Gamma}^a\left(0; \theta\right)$ in \eqref{eq:GCov22} with  $diag\left(\hat{\Gamma}_d(0; \theta)\right)$ containing only the diagonal elements of $\hat{\Gamma}_a(0; \theta)$. This latter version of the GCov estimator is no longer semiparametrically efficient, as it is not optimally weighted [see \cite{gourieroux2023generalized}, \cite{cubadda2011testing}]. Another strategy for handling problems with the inversion of matrix $\hat{\Gamma}^a\left(0; \btheta\right)$  in high-dimensional settings is to replace the GCov estimator with
the regularized RGCov estimator proposed by \cite{giancaterini2025regularized}, which can preserve the semiparametric efficiency under suitable conditions.}.
% the diagonal GCov 

The choice of an informative set of transformations ($a_{k}, k=1,..., K$) depends on the specific time series under investigation. For example, in financial applications, linear and quadratic functions can be selected, such as $a_1(\epsilon_t) =\epsilon_{t}$, $a_2(\epsilon_t) = \epsilon_{t}^2$. This implies that $a_1$ is a linear function of errors in a causal-noncausal process, while $a_2$ transforms the error term by squaring it for each $t=1, \dots, T$. In application to MAR processes with heavy-tailed error distributions, such as $\alpha$-stable distributions, including Cauchy, or t-student distribution with low degrees of freedom, one can use the square root of absolute values, or other fractional powers to ensure the existence of moments of the transformed errors and the asymptotic normality of the GCov estimator. 

Moreover, the objective function minimized in \eqref{eq:GCov22} and evaluated at the estimated parameter $\hat{\theta}_T$ can be used to test the fit of the model. Specifically, we test the null hypothesis $H_0 : \{\Gamma^a_0 (h) =0, \; h=1,...,H\},$ using the statistic 
\begin{equation*}
    \hat{\xi}_T(H) =T L_T( \hat{\theta}_T,H),
\end{equation*}
\nin which, under the implicit null hypothesis of serial independence of errors, has an asymptotic chi-square distribution with degrees of freedom equal to $H K^2 -dim(\theta)$  [\cite{gourieroux2023generalized}]. 

\section{Bubble Analysis}

Our approach to bubble analysis assumes that the process of interest follows a strictly stationary MAR($r,s$) process with a heavy-tailed error distribution. As mentioned earlier, among the heavy-tailed distributions are the $\alpha$-stable distributions  
including the Cauchy distribution 
and t-student distributions with low degrees of freedom.
In this Section, we focus our attention on MAR processes of orders $r$ and $s$ such that the combined order $p=r+s$ is less than or equal to 2, so that the process of interest is either a MAR(0,1), i.e., a purely noncausal process of order 1, or a MAR(1,1), or a causal MAR(1,0). These processes are geometrically ergodic. \cite{gourieroux2017local} and \cite{fries2019mixed} also show that the MAR(0,1) and MAR(1,1) processes are Markov of order 1 and 2, respectively. The results can be generalized to higher-order MAR($r,s$) processes, which are  Markov too [\cite{fries2019mixed}, Proposition 3.1].

\subsection{On-Bubble Dynamics}

Consider the locally explosive MAR(0,1) and MAR(1,1) processes. It has been shown in the literature that during a bubble episode, and conditional on $y_t >y$, where $y$ is large, the causal-noncausal MAR(0,1) and MAR(1,1) processes with $\alpha$-stable distributed errors 
have a distinct dynamic [\cite{fries2022conditional}], which leads to different behavior of the conditional autocovariances of its latent components. In this Section, we describe these dynamics and generalize the results in \cite{fries2022conditional}  using the concept of a spectral tail process, based only on observations above a high threshold $y$. Our results are more general than those in \cite{fries2022conditional} or \cite{de2025forecasting} as we do not need to assume a distribution such as the $\alpha$-stable for the whole distribution, but only conditions on the tail process. We consequently define a bubble in this paper as follows:
\medskip

\nin \textbf{ Definition:} The bubble occurs when the observations exceeding a high threshold from a strictly stationary autoregressive noncausal process with i.i.d. errors and a heavy-tailed error distribution with tail index $\alpha$ increase (decline) approximately at the rate of a spectral tail process.

\medskip

\nin The spectral tail process captures and measures the extremal dependence, which determines the shape and duration of a bubble. It is characterized as follows:

\medskip
\nin \textbf{Proposition 1} [\cite{kulik2020heavy}, Chapter 15.3]:
\nin Let $\{y_t, t \in \mathbb{Z}\}$ be a strictly stationary process with i.i.d. errors $\epsilon_t,  t=1,2,...$ and a heavy-tailed error distribution with (Pareto-type) tails of tail index $\alpha$ and with a  two-sided MA representation:

$$y_t =  \sum_{h=-\infty}^{+ \infty} c_h \epsilon_{t-h},$$

\nin with nonnegative coefficients. The process  $\left( \Frac{y_{t+h}}{y_t}\right)_h, h \in \mathbb{Z}$, converges in distribution to the (spectral) tail process $(X_h)$, so that conditional on a large $y_t>y$, where $y$ is a high threshold, we have :
$${\cal L} \left( \frac{y_{t+h}}{y_t},\; h=-H,...,H| y_t>y \right) \stackrel{d}{\rightarrow}{\cal L} (X_h, \; h=-H,...,H),$$

\nin where 
$\stackrel{d}{\rightarrow}$ denotes convergence in distribution and ${\cal L}$ stands for "law". 

It is natural to interpret this convergence
as an approximation $\Frac{y_{t+h}}{y_t} \stackrel{d}{\approx} X_h$ for $h \in \mathbb{Z}$ [\cite{drees2020peak}], where
\begin{equation}
X_h = \frac{c_{h+N}}{c_N} X_0, \; h \in \mathbb{Z},
\end{equation}

\nin and where $X_0=1$, and $N$ is an integer-valued random variable, independent of $y$ and such that: 

\begin{equation}
 P[N=h] = \frac{c_h^{\alpha}}{\sum_{h \in \mathbb{Z}} c_h^{\alpha}}, \;\; \forall h \in \mathbb{Z} .
\end{equation}

\medskip

\nin  By setting $X_0=1$, we consider an upward (positive) bubble. Henceforth, we focus our attention on this case for ease of exposition as well as for the pattern of the bubbles observed on the "green" indicators investigated in this paper.
Similar results are easily obtained for downward bubbles, conditional on $y_t<y$ where $y$ tends to $-\infty$ with $X_0=-1$.

Proposition 1 can be applied to the MAR(1,1) and MAR(0,1) processes as follows.
Let us first consider the MAR(1,1) process defined in Section 2.1, assuming positive coefficients $\phi$ and $\psi$. 

\medskip

\nin \textbf{Proposition 2:}
The strictly stationary MAR(1,1) process with i.i.d. errors $\epsilon_t, t=1,2,...$ and a heavy-tailed error distribution with tail index $\alpha$:
$$(1 - \phi L)(1-\psi L^{-1})y_t  = \epsilon_t,$$

\nin (i) admits a two-sided MA($\infty$) representation with the coefficients:
%\begin{equation}
$c_h = \Frac{1}{1-\phi \psi} \psi^{-h}$, if $ h \leq 0$, \;\;\mbox{and} \;\;$c_h = \Frac{1}{1-\phi \psi} \phi^{h}$, if $ h \geq 0$.
%\end{equation}

\nin (ii) the tail process $ X_h$ is such that:
$$X_h = \psi^{-1} X_{h-1} \id_{N \leq - h} + \phi X_{h-1} \id_{N > - h},$$

\nin with $X_0=1$, and the probability $ P[N=h]$:

\medskip

\nin $P[N=h] =  \Frac{\psi^{-h \alpha}}{[\frac{1}{1-\phi^{\alpha}}+ \frac{1}{1-\psi^{\alpha}} -1]} $, if $ h \leq 0$, and
$P[N=h] = \Frac{\phi^{h \alpha}}{[\frac{1}{1-\phi^{\alpha}}+ \frac{1}{1-\psi^{\alpha}} -1]} $, if $ h \geq 0$,

\medskip
\nin with $0^0 = 1$, by convention.

\medskip
\nin \textbf{Proof:} See Appendices A1-A2. 

\medskip 

\nin The random variable $N$ determines the time to the peak of a bubble and its overall duration.
We observe that the distribution of variable $N$  is a mixture of two geometric distributions (see Proposition 4).

\medskip 
\nin \textbf{Corollary 1:}
In the MAR(1,1) process, we have:
$$X_h = \phi^h \id_{N > Max(-h,0)} + \phi^{-h-N} \psi^{-N} \id_{0<N \leq -h} + \phi^{h+N} \psi^{N}
\id_{-h <N \leq 0} + \psi^{-h}  \id_{N \leq Min(-h,0)}$$

During a bubble episode, we can distinguish the phases of growth and decline.
The noncausal persistence of MAR(1,1) processes determines the rate at which a bubble keeps growing up to time $t-N$. 
The negative power of $\psi$
indicates that the growth is explosive while $N<0$ given that $|\psi|<1$ is assumed for strict stationarity. The above result is consistent with Proposition 4.2 in \cite{fries2022conditional} which describes the dynamics of a MAR(1,1) process conditional on a large value $y_t>y$ and $\frac{y_{t+1}}{y_t} = \frac{y_{t+2}}{y_{t+1}} = 1/\psi$. In the MAR(1,1), conditional on  $y_t>y$, the bubble either keeps growing  to the next value $y_{t+1} = \frac{1}{\psi} y_t$, or it bursts and decreases  to $y_{t+1} = \phi y_t$.
This latter pattern is also consistent with formula (v) given in the proof of Corollary 1, Appendix A.2a, for $N > 0$, which additionally implies that the bubble bursts at $N=0$.

In a MAR(0,1), the bubble bursts vertically to 0, while in the causal AR(1), i.e., MAR(1,0), we observe a jump, followed by a decline determined by the autoregressive coefficient, as shown below.

\medskip

\nin \textbf{Corollary 2:}
The strictly stationary MAR(0,1) process with i.i.d. errors $\epsilon_t, t=1,2,,..$ and a heavy-tailed error distribution with tail index $\alpha$:
$$(1-\psi L^{-1}) y_t = y_t - \psi y_{t+1} = \epsilon_t,$$

\nin admits a one sided  MA($\infty$) representation with the moving average coefficients:

$$
c_h  =  \psi^{-h},\;\;\mbox{if}\; h \leq 0, \;\;\mbox{and} \;\;
c_h  =  0, \;\; \mbox{if} \; h > 0.
$$

\nin The tail process is such that:
$$X_h = \psi^{-1} X_{h-1} \id_{N \leq -h},$$

\medskip

\nin with $X_0=1$ and the probability 
$ P[N=h]$:

\medskip
\nin $P[N=h] = 0$,  if $ h > 0$,
 and $P[N=h]= (1-\psi^{\alpha}) \psi^{-h \alpha} $, if $ h \leq 0$,

\medskip

\nin Because the bubble bursts vertically in a MAR(0,1) process, variable $N$ takes only negative values in that process. It follows from Proposition 2 that the variable $-N$ has a geometric distribution on 
$\mathbb{N}$ with parameter $\psi^{\alpha}$ during the growth phase of bubble.

\bigskip

\nin The causal autoregressive processes with i.i.d. errors and a heavy-tailed error distribution with tail index $\alpha$ may admit jumps followed by tail process behavior.

\medskip

\nin \textbf{Corollary 3:} The strictly autoregressive of order 1 (causal AR(1), i.e. MAR(1,0)) process:
$$y_t  =  \phi y_{t-1}  + \epsilon_t, $$

\nin with i.i.d. errors $\epsilon_t, t=1,2,,..$ and a heavy-tailed error distribution with tail index $\alpha$ and $0<\phi <1$ 
admits a one sided MA($\infty$) with coefficients:

$$
c_h  =  \phi^{h}\;\;\mbox{if}\; h \geq 0, \;\;\mbox{and} \;\;
c_h  =  0, \;\; \mbox{if} \; h < 0.
$$

\nin The tail process is such that:
$$X_h = \phi X_{h-1} \id_{N \geq h},$$

\nin with $X_0=1$ and the probability $ P[N=h]$ :

\nin $P[N=h] = 0$;  if $ h < 0$, and $P[N=h] = (1-\phi^{\alpha}) \phi^{h \alpha} $; if $ h \geq 0$,

\medskip

\nin Hence, the path of $y_t$ displays jumps, with a geometric decline. The variable $N$ has a geometric distribution with parameter $\phi^{\alpha}$ during the decline following a jump. Moreover, the behavior of pure noncausal autoregressive processes of order 2 (MAR(0,2) with i.i.d. errors $\epsilon_t, t=1,2,..$ and a heavy-tailed error distribution with tail index $\alpha$  is described in Appendix B.

\subsection{Tail Behavior of Latent Components}

\nin The latent components of a MAR(1,1) process are defined in equations (3) and (4) as:
$$u_{t+1} = (1-\phi L) y_{t+1} = y_{t+1} - \phi y_{t}\; \mbox{and} \;v_t = (1-\psi L^{-1}) y_t = y_t -\psi y_{t+1}, t=1,2,...$$

\nin We observe that the ratios of these latent components divided by $y_t$:
$$\Frac{u_{t+h}}{y_{t}} = \Frac{y_{t+h} - \phi y_{t+h-1}}{y_{t}} \;\mbox{and}\; \Frac{v_{t+h}}{y_{t}} = \Frac{y_{t+h} - \psi y_{t+h+1}}{y_{t}}$$

\nin are functions of ratios $y_{t+h}/y_t$ for $h \in -H,..,H$. From Proposition 1, it follows that at any time $t$ when $y_t>y$ where $y$ is a high threshold, the ratios of observations $y_{t+h}/y_t \stackrel{d}{\approx} X_h$ for any $h$, where $\stackrel{d}{\approx}$ denotes approximately equal to in distribution. Our goal is to replace the above ratios by the tail components: $U_h = X_{h} - \phi X_{h-1}$ and $ V_h= X_h- \psi X_{h+1}$, respectively. Then, when $y_t>y$ is large, the above ratios become linear deterministic functions of the components $X_h$ of the tail process.

Let us now examine how Proposition 2 can be used to obtain the asymptotic behavior (in distribution) of  $U_h, V_h, \; h = -H,..., H$. 

\medskip
\nin \textbf{ Proposition 3:} At time $t$ such that $y_t>y$, for a high threshold $y$, we have:
$$u_{t+h}/y_t \stackrel{d}{\approx} X_h - \phi X_{h-1} := U_h, \; v_{t+h}/y_t \stackrel{d}{\approx} X_h - \psi X_{h+1} := V_h,$$
\nin and
$$U_h = (\psi^{-1} - \phi) X_{h-1} \id_{N \leq -h}, \; V_h = (1-\psi \phi) X_{h} \id_{N > -h-1}.$$

\nin \textbf{Proof:} Let us consider the causal component. It follows from Proposition 2 that:
$$U_h = X_h - \phi X_{h-1} = (\psi^{-1} - \phi) X_{h-1} \id_{N \leq -h}.$$
Similarly, we have:
$$V_h = X_h - \psi X_{h+1} =  X_h - X_h \id_{N \leq -h-1}  - \psi \phi X_h \id_{N > -h-1} = (1-\phi \psi) X_h \id_{N > -h-1}.$$

\medskip
\nin Let us now examine the values of $U_h$ and $V_h$ during a bubble episode. We can show that either one of them is zero during a bubble episode as follows:

\medskip
\nin \textbf{Corollary 4:} The causal and noncausal tail components $U_h, V_h$ are such that:

\medskip

$U_h = 0$, if $N>-h <=> t+h > t-N$,

$V_h = 0$, if $N\leq-h-1 <=> t+h < t-N-1$,

\medskip

\nin Proposition 3 and Corollary 4 can be interpreted as follows. Let us consider an exogenous time $t$ when $y_t>y$, for large $y$. Then $t$ belongs to a bubble episode, with a peak at time $t-N$. The noncausal (resp. causal) tail component is equal to zero after (resp. before) the peak and has the behavior of a MAR(0,1) tail process before the peak (resp. MAR(1,0) tail process after the peak). This result can also be used to determine the behavior of the products of the causal and noncausal tail components. As shown below in Corollary 5, since during the bubble episode either $U_h$ or $V_h$ is zero, their product is zero during the entire bubble episode.

\medskip

\nin \textbf{Corollary 5:}  We have
$$\xi_{t,h,k} = \frac{u_{t+h}v_{t+k}}{y_{t}^2} \stackrel{d}{\approx} U_h V_k =0, \;\; \mbox{if} \;\; h-k \geq 1.$$

\nin \textbf{Proof}: We see that $U_hV_k = 0 \iff (N >-h)$ or $(N \leq -k-1)$.
This condition is equivalent to $N \in \mathbb{Z}$ that is always satisfied iff:
\begin{eqnarray*}
\mathbb{Z} & = & (-\infty, -k-1] \cup [-h+1, \infty]  \\
& \iff & - h +1 \leq -k -1 +1 =-k \\
& \iff & h-k \geq 1.
\end{eqnarray*}

\nin The result follows. In particular, we have:
$$\xi_{t,h}= \Frac{u_{t+h+1}v_{t+h}}{y_{t}^p} \stackrel{d} {\approx} U_{h+1} V_h = 0, \; \forall h$$ 

\nin where $p=r+s=2$ is the combined autoregressive order of the process, and
$$\xi_{t,0}= \Frac{u_{t+1}v_t}{y_t^p} \stackrel{d}{\approx} U_1 V_0 = 0.$$

\nin In particular, for the MAR(0,1) process with $p=1$, we get: 

(i) at $h=0$. 
$$
\xi_{t,0} = \frac{v_{t}}{y_t} \stackrel{d}{\approx} 1-  \psi X_1,
$$
\nin where
$$X_1 = \psi^{-1} \id_{N \leq -1}. $$

\nin Then, $\xi_{t,0} = 1 - \id_{N \leq -1}$ and is approximately always 0 during the bubble.

(ii) at $h=1$. 
$$
\xi_{t,1} = \frac{v_{t+1}}{y_t} \stackrel{d}{\approx} X_1-\psi X_2 
$$
\nin where
$$X_2 = \psi^{-1} X_1 \id_{N \leq -2} $$

\nin Then, $\xi_{t,1}$ is approximately equal to 0 as well.

\nin The observed behavior of $\xi_{t,h}$ can be used for testing the hypothesis that the process is in a bubble episode at time $t+h$. This approach is pursued in Section 4. Below, we discuss the behavior of the variable $N$ and associated inference.

\subsection{Duration and Time to Peak of a Bubble}

This section describes the stochastic properties of the random variable $N$, which determines the duration of a bubble, and of $t-N$ for $N\leq 0$, which is interpreted as the time to peak. We consider first the MAR(1,1) process.

\medskip

\nin \textbf{Proposition 4:}
In the MAR(1,1) process, the distribution of $N$ is a mixture of two geometric distributions with $P[N \geq 0] = \Frac{1}{1-\phi^{\alpha}}/(\frac{1}{1-\phi^{\alpha}} + \frac{\psi^{\alpha}}{1-\psi^{\alpha}})$. The distribution of $N$ given $N \geq 0$ is a geometric distribution in $\mathbb{N}$ with parameter $\phi^{\alpha}$, and the distribution of $-1-N$, given $N \leq -1$ is a geometric distribution on $\mathbb{N}$ with parameter $\psi^{\alpha}$.

\medskip
\nin \textbf{Proof:} From Proposition 2, it follows that:
$$ P[N \geq 0] = \Frac{1}{1-\phi^{\alpha}} / \left( \frac{1}{1-\phi^{\alpha}} +
\frac{\psi^{\alpha}}{1-\psi^{\alpha}} \right).$$

\medskip

\nin  Then, for $h \geq 0$:

\nin $P(N=h| N \geq 0) = \Frac{\phi^{h \alpha}}{1-\phi^{\alpha}}$, which is a geometric distribution on $\mathbb{N}$ with the parameter $\phi^{\alpha}$.

\medskip

\nin For $h <0$, we have:

\nin $P(N=h| N < 0) = \Frac{\psi^{(-h-1)\alpha}}{1-\psi^{\alpha}}$, which 
means that $-1-N$
has a geometric distribution on  $\mathbb{N}$ with parameter $\psi^{\alpha}$.

\medskip
Proposition 4 allows us to infer about the bubble duration from the moments of variable $N$. In practice, one may be interested in finding the marginal expected value of $N$ as it is informative about the shape of a bubble and can also reveal an asymmetry in the growth and burst phases.

\medskip

\nin \textbf{Corollary 6:}
\nin The expected value of $N$ is: 
$$E(N) = \Frac{1}{\frac{1}{1-\phi^{\alpha}} + \frac{1}{1-\psi^{\alpha}} -1}
\left( \Frac{\phi^{\alpha}}{(1-\phi^{\alpha})^2} - \Frac{\psi^{\alpha}}{(1-\psi^{\alpha})^2} \right).$$

\nin \textbf{Proof:} We have
\begin{eqnarray*}
E(N) & = & E(N|N \geq 0) P(N \geq 0) + E(N|N \leq -1) P(N \leq -1) \\
& = & E(Z_1)P(N \geq 0) + E(-1-Z_2) P(N \leq -1), 
\end{eqnarray*}

\nin where $Z_1, Z_2$ follow geometric distributions on $\mathbb{N}$ with parameters $\phi^{\alpha}$ and $\psi^{\alpha}$, respectively. It follows that:
\begin{eqnarray*}
E(N) & = & \Frac{1}{\frac{1}{1-\phi^{\alpha}} + \frac{\psi^{\alpha}}{1-\psi^{\alpha}}} 
\left[  \Frac{1}{1-\phi^{\alpha}} \frac{\phi^{\alpha}}{1-\phi^{\alpha}} + \frac{\psi^{\alpha}}{1-\psi^{\alpha}} (-1- \frac{\psi^{\alpha}}{1-\psi^{\alpha}}) \right] \\
& = & \Frac{1}{\frac{1}{1-\phi^{\alpha}} + \frac{\psi^{\alpha}}{1-\psi^{\alpha}}} \left( \Frac{\phi^{\alpha}}{(1-\phi^{\alpha})^2} - \Frac{\psi^{\alpha}}{(1-\psi^{\alpha})^2} \right). 
\end{eqnarray*}

\begin{figure}
    \centering
    \caption{$E(N)$ for MAR(0,1) and MAR(1,1)}
    \begin{subfigure}[t]{0.48\textwidth}
            \includegraphics[width=\linewidth]{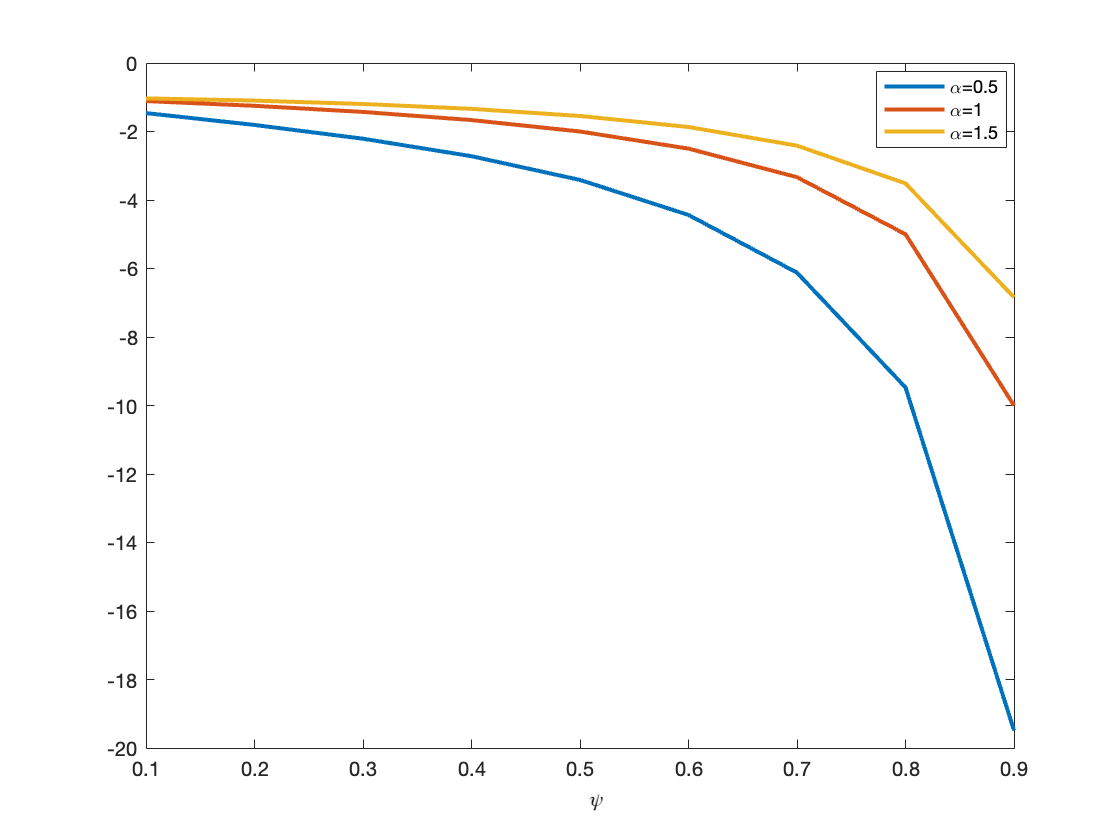}
            \caption{$E(N)$ for MAR(0,1)}
            \label{EN01}
    \end{subfigure}
        \hfill
    \begin{subfigure}[t]{0.48\textwidth}
            \includegraphics[width=\linewidth]{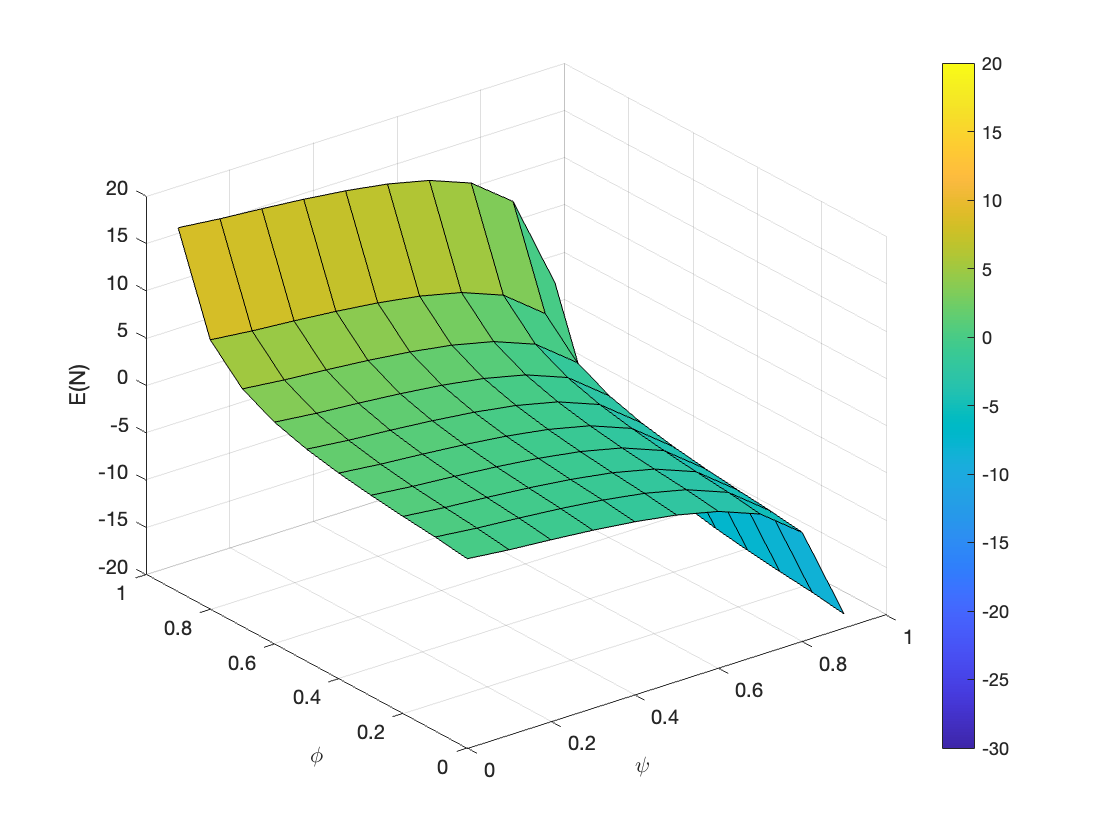}
            \caption{$E(N)$ for MAR(1,1), $\alpha=0.5$}
            \label{EN1105}
    \end{subfigure}
    \hfill
        \begin{subfigure}[t]{0.48\textwidth}
            \includegraphics[width=\linewidth]{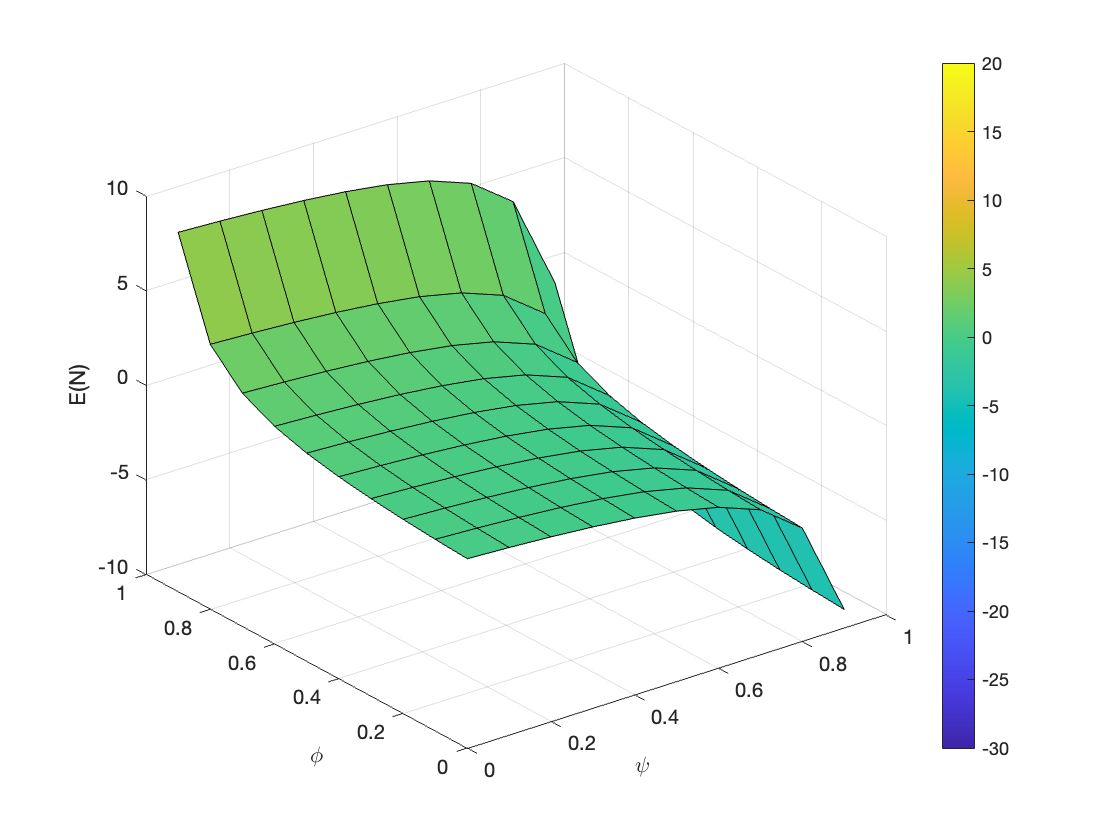}
            \caption{$E(N)$ for MAR(1,1), $\alpha=1$}
            \label{EN111}
    \end{subfigure}
        \hfill
    \begin{subfigure}[t]{0.48\textwidth}
            \includegraphics[width=\linewidth]{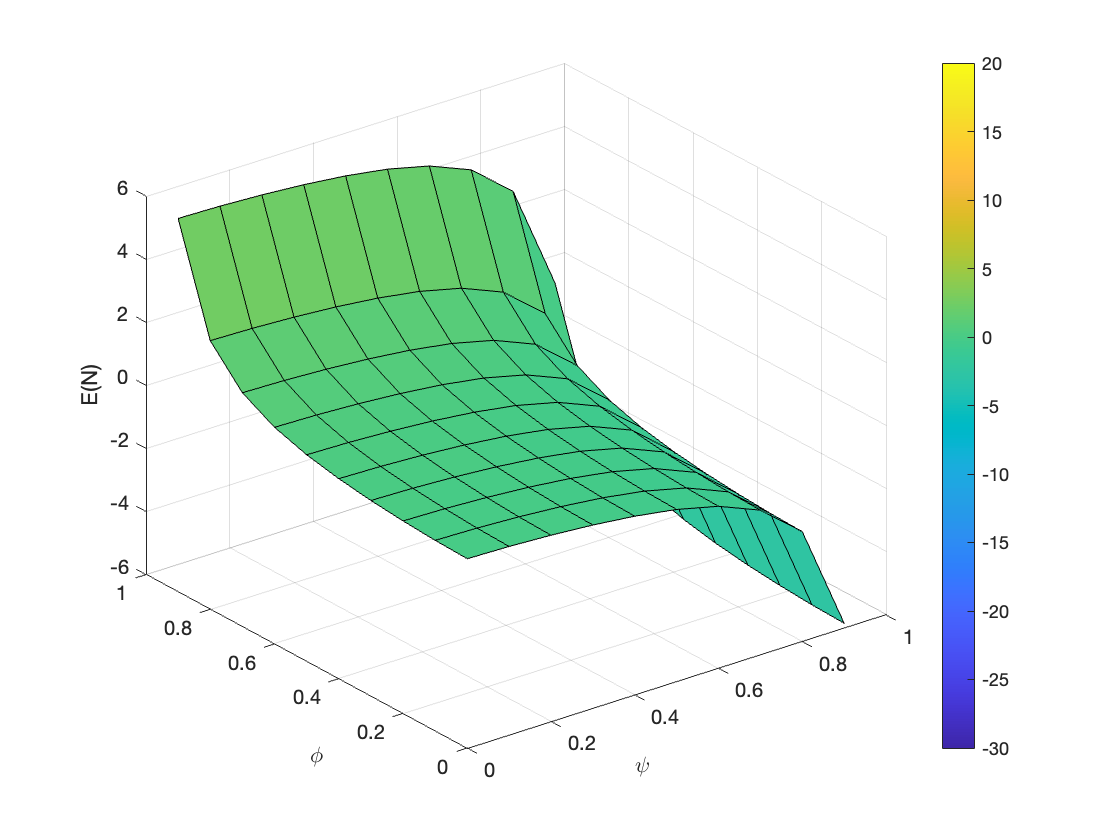}
            \caption{$E(N)$ for MAR(1,1), $\alpha=1.5$}
            \label{EN1115}
    \end{subfigure}
    \label{EN}
\end{figure}

\nin The marginal expected value of $N$ depends on the values of $\phi$ and $\psi$ and is symmetric in $\phi$ and $\psi$. In particular, it is equal to $0$ if $\phi=\psi$, which corresponds to a symmetric bubble. In addition, this expectation depends on $\alpha$.  The effects of $\phi, \psi, \alpha$  are illustrated in Figure \ref{EN}, where we observe that, for a fixed $\psi$, the expected value $E(N)$ increases in $\phi$. For a fixed $\phi$, $E(N)$ increases in absolute values in $\psi$. In addition, the range of values of $E(N)$ depends on the tail parameter $\alpha$.

If the bubble is asymmetric, then $E(N)$ can take either a positive or negative value. Therefore, one may be interested in computing the conditional expectation of $N$ given $N<0$ to approximate the expected time to peak. 

It is also interesting to compute the cumulative distribution function (cdf), or survival function of $N$, to derive the quantiles of the distribution of $N$ and obtain a prediction interval for $N$.  We find the quantiles $h^*_U$ and $h^*_L$ of the distribution of $N$ at levels $(1-\gamma)$ and $\gamma$, where $\gamma$ is small, which are: $h^*_U \;\; \mbox{such that} \;\; P[N \leq h^*_U] = \gamma, \;\mbox{and} \;\; h^*_L \;\; \mbox{such that} \;\; P[N>h^*_L] = 1-\gamma$. Using small $\gamma$, we separately consider the geometric distributions of the mixture.

\medskip
\nin \textbf{Corollary 7:} For $h \leq 0$, sufficiently small, we have:
$$
 P[N \leq h] 
= \Frac{\psi^{-h \alpha}}{1-\psi^{\alpha}} \Frac{1}{[\frac{1}{1-\phi^{\alpha}} +
\frac{1}{1-\psi^{\alpha}} -1]}, 
$$

\nin and for $h>0$, sufficiently large, we have:
$$P[N>h] = \Frac{\phi^{h \alpha}}{1-\phi^{\alpha}} \Frac{1}{[\frac{1}{1-\phi^{\alpha}} +
\frac{1}{1-\psi^{\alpha}} -1]}. $$

\nin \textbf{Proof:} see Appendix.

\medskip

\nin The expected value (median or mode) provides a point prediction of the variable $N$. The quantiles of the distribution
of $N$ provide the prediction interval for $N$.

\medskip

\nin Let us now consider the pure noncausal MAR(0,1) process characterized by a vertical bubble burst. For the MAR(0,1) process, the variable $-N$ follows a geometric distribution on $\mathbb{N}$ with parameter $\psi^{\alpha}$ (by Corollary 2 or Proposition 4) during bubble growth. The cumulative probability function of $N$ is:

$$P[N \leq h] = (1-\psi^{\alpha}) \sum_{i \leq h} \psi^{-i \alpha} =(1-\psi^{\alpha}) \sum_{i \geq -h} \psi^{i \alpha} = \psi^{-h \alpha},$$

\nin for $h \leq 0$. The mean of this geometric distribution  is equal to 
$$E(N) = - \Frac{1}{1-\psi^{\alpha}},$$

\nin and can be interpreted as the average time to peak. The median is:
$$med(N) =\left[ \Frac{\log 2}{ \alpha \log (\psi)}\right],$$

\nin where the brackets denote the integer part, and the mode is denoted by $mode(N)=0$. Since the distribution of $-N$ is skewed, the expectation, median, and mode provide three different point predictions. We illustrate the effect of the parameters $\psi$ and $\alpha$ in Figure \ref{EN01}. We observe that for MAR(0,1), the absolute value of the expected time to peak increases in $\psi$. This is consistent with the dynamics of this process, with a growing phase of a bubble followed by an instantaneous crash. Moreover, the values of $E(N)$ depend on the parameter $\alpha$.

 Up to the effect of the integer part, both the median and the expectation of $N$ increase in the absolute value of $\psi^{\alpha}$. To build the prediction interval, we use the quantiles of the geometric distribution, which are determined by inverting the cumulative probability function $P[N \leq h] = \psi^{-h \alpha}$. Given that the quantile at level $\gamma$ is equal to :
$$h(\gamma) = - \Frac{\log \gamma}{\alpha \log \psi},$$

\nin the prediction interval for $N$ at level 10\% is:
$$(- \Frac{\log 0.05}{\alpha \log \psi}, - \Frac{\log 0.95}{\alpha \log \psi} ).$$

\nin Because the variable is discrete, we can round up or down the upper and lower quantiles to integer values to make it more conservative. 

In practice, the predictions and prediction intervals of $N$ can be found by replacing the unknown parameters $\phi, \psi$ by their GCov-based estimates based on a sample of $T$ observations, which are also used for computing the test statistics given in the next section. The tail parameter $\alpha$ can be approximated by the Hill estimator [see e.g. eq. (9.5.1), Section 9.5, \cite{kulik2020heavy}], which suffers from the bias-variance trade-off in finite samples. \cite{kulik2020heavy} show that under the tail process approach, the Hill estimator is consistent and asymptotically normally distributed.

\section{Inference}

\subsection{The Statistics}

Let us show how the results derived in Section 3.2 can be used to build diagnostic tools to test the MAR(1,1) process for bubbles. Diagnostics are based on the counterparts of 
$$\xi_{t,0} = \Frac{u_{t+1} v_{t}}{y_t^2} = \left(\frac{y_{t+1}}{y_t} -\phi_0 \right) 
\left(1-\psi_0 \frac{y_{t+1}}{y_t} \right),$$

\nin where $\phi_0, \psi_0$ denote the true values of the parameters for a given observed process, we wish to estimate from a sample of length $T$. The quantity $\xi_{t,0}$ depends on the observations and the unknown parameters. The unknown parameters can be replaced by consistent and asymptotically normally distributed estimators $\hat{\phi}_T, \hat{\psi}_T$ of $\phi,\psi$ obtained from a sample of $T$ observations. Then, the sample counterpart of $\xi_{t,0}$ is:
$$\hat{\xi}_{t,T}(0) = \left( \frac{y_{t+1}}{y_{t}}  - \hat{\phi}_T \right) \left(1- \hat{\psi}_T \frac{y_{t+1}}{y_t} \right).$$

\nin The statistics $\hat{\xi}_{t,T}(0), t=1,...,T-1$ can be used as follows. From Corollary 5, we know that if  $y_t$ is sufficiently large at time $t$, then $\xi_{t,0} \stackrel{d}{\approx} U_1 V_0$ and that $U_1 V_0 = 0$.
Therefore, we expect that $\hat{\xi}_{t,T}(0)$ is close to 0. We can also consider another lag $h$ with: 
$$
\xi_{t,h} =   \Frac{u_{t+h+1} v_{t+h}}{y_t^2} = \left(\frac{y_{t+h+1}}{y_t} -\phi_0  \frac{y_{t+h}}{y_t} \right) \left(\frac{y_{t+h}}{y_t} -\psi_0 \frac{y_{t+h+1}}{y_t} \right). $$

\nin  It was shown in Corollary 5 that  $\xi_{t,h} \stackrel{d}{\approx} U_{h+1} V_h$ and $U_{h+1} V_h = 0$. Its sample counterpart is
$$
\hat{\xi}_{t,T}(h) =   \Frac{\hat{u}_{t+h+1} \hat{v}_{t+h}}{y_t^2} = \left(\frac{y_{t+h+1}}{y_t} - \hat{\phi}_T  \frac{y_{t+h}}{y_t} \right) \left(\frac{y_{t+h}}{y_t} - \hat{\psi}_T \frac{y_{t+h+1}}{y_t} \right), $$

\nin which can be used for inference at higher horizons.

\subsection{Confidence Band}

Consider the statistic $\hat{\xi}_{t,T}(0)$. The zero value of the transformed tail process can be considered as the true value of some tail parameter $\theta_1 = U_1 V_0$, which is deterministic and equal to 0 if $y_t$ is sufficiently large, and it is stochastic, otherwise. Then, at each exogenous time $t$, we can consider the null hypothesis:
$$H_{0,1} = \{ \theta_1 = 0\}.$$

\nin The difficulty is that we have a double asymptotic in the level of threshold $y$ and in the number of observations $T$. To derive reliable confidence bands, we assume that during the bubble episode,
the uncertainty in the approximation $\xi_{t,0} \stackrel{d}{\approx} U_1V_0$ is negligible with respect to the asymptotics in $T$.
Then, during a bubble episode when $H_{0,1}$ is satisfied and $y_{t} >y$, with large $y$, we have conditional on $y_t, y_{t+1}$:
\begin{eqnarray*}
& & \sqrt{T} ( \hat{\xi}_{t,T}(0) - \xi_{t,0}) \stackrel{d}{=} \sqrt{T} (   \hat{\xi}_{t,T}(0) - U_1V_0) \\
& = & \sqrt{T}  \hat{\xi}_{t,T}(0) \stackrel{d}{\rightarrow} N(0, \sigma_t^2),
\end{eqnarray*}

\nin where $ \sigma_t^2$ is obtained by the delta method as follows: 

$$ \sigma_t^2 = V_t \left\{ \left[ -\left( 1 - \psi_0 \frac{y_{t+1}}{y_{t}} \right),
\frac{-y_{t+1}}{y_t} \left(  \frac{y_{t+1}}{y_{t}} - \phi_0   \right) \right] \sqrt{T} \left( \begin{array}{c} \hat{\phi}_T - \phi_0 \\ \hat{\psi}_T - \psi_0 \end{array} \right) \right\}.$$

\nin This quantity is then consistently estimated from:
$$ \hat{\sigma}_{t,T}^2(0) = \left( \frac{\hat{v}_t}{y_t}, \frac{y_{t+1} \hat{u}_{t+1}}{y_{t}^2} \right) \hat{\Omega}_T \left( \begin{array}{c} \Frac{\hat{v}_t}{y_t} \\ \Frac{y_{t+1} \hat{u}_{t+1}}{y_{t}^2} 
\end{array}   \right),$$

\nin where $\hat{\Omega}_T$ is a consistent estimator of the asymptotic variance matrix of the parameter estimators. Then, under the assumption of a MAR(1,1) process and  following a large $y_t$, we have 
$$| \sqrt{T} \, \hat{\xi}_{t,T}(0)/\hat{\sigma}_{t,T}(0)| \leq 1.96, $$

\nin with an asymptotic probability of 95$\%$. This leads to a functional diagnostic tool that consists of reporting for any time $t$ the quantities $\hat{\xi}_{t, T}, \; t=1,..., T-1$ along with the band

$$\left(\hat{\xi}_{t,T}(0) \pm 1.96 \frac{\hat{\sigma}_{t,T}(0)}{\sqrt{T}} \right).$$

Since the tail process does not depend on the level of $y_t$ and the values of $y_{t+1}$, the width of the band is independent of time $t$, $t=1,..., T$. Then, the times at which the statistic is inside the band are the times associated with a bubble with probability 95\%.

\medskip
This approach is easily extended to other lags. From the statistic $\hat{\xi}_{t,T}(h)$, we can test the null hypothesis:
$$H_{0,h} = \{ \theta_h = 0\}$$

\nin where $\theta_h = U_{h+1}V_h$ and $y_{t} >y$. Then, conditional on $y_t, y_{t+h}, y_{t+h+1}$ we have:
\begin{eqnarray*}
& & \sqrt{T} ( \hat{\xi}_{t,T}(h) - \xi_{t,h}) \stackrel{d}{=} \sqrt{T} (   \hat{\xi}_{t,T}(h) - U_{h+1}V_h) \\
& = & \sqrt{T}  \hat{\xi}_{t,T}(h) \stackrel{d}{\rightarrow} N(0, \sigma_t^2(h)) 
\end{eqnarray*}

\nin where $ \sigma_t^2(h)$ is: 
$$ \sigma_t^2(h) = V_t \left\{ \left[ \frac{-y_{t+h}}{y_{t}} \left( \frac{y_{t+h}}{y_t}- \psi_0 \frac{y_{t+h+1}}{y_t} \right),
\frac{-y_{t+h+1}}{y_{t}} \left(\frac{y_{t+h+1}}{y_{t}} - \phi_0 \frac{y_{t+h}}{y_{t}}  \right) \right] \sqrt{T} \left( \begin{array}{c} \hat{\phi}_T - \phi_0 \\ \hat{\psi}_T - \psi_0 \end{array} \right) \right\}.$$

\nin This quantity is then consistently estimated from:
$$ \hat{\sigma}_{t,T}^2(h) = \left( \frac{y_{t+h} \hat{v}_{t+h}}{y_{t}^2}, \frac{y_{t+h+1} \hat{u}_{t+h+1}}{y_{t}^2} \right) \hat{\Omega}_T \left( \begin{array}{c} \Frac{y_{t+h} \hat{v}_{t+h}}{y_{t}^2} \\ \Frac{y_{t+h+1} \hat{u}_{t+h+1}}{y_{t}^2} 
\end{array}   \right).$$
\nin Under the assumption of a MAR(1,1) process and following a large $y_{t}>y$:
$$| \sqrt{T} \, \hat{\xi}_{t,T}(h)/\hat{\sigma}_{t,T}(h)| \leq 1.96, $$

\nin with the asymptotic probability of 95 \%. This leads to a set of functional diagnostic tools, where for any time $t$ and $h$ the quantities $\hat{\xi}_{t,T}(h), \; t=1,...,T-h-1$ are reported along with the band:

$$\left(\hat{\xi}_{t,T}(h) \pm 1.96 \frac{\hat{\sigma}_{t,T}(h)}{\sqrt{T}} \right).$$

\medskip
In practice, the above procedure may not distinguish between a bubble and a short-lasting spike with the same growth rate. In addition, there can be times $t$ during the bubble when the statistic is outside the band because either (i) the MAR(1,1) model is misspecified, or (ii) the MAR(1,1) is well specified, but the value of $y_t$ is not sufficiently large.

This method easily provides the test statistics for the MAR(0,1) and MAR(1,0) processes, based on Corollary 5, with $\hat{\xi}_{t,T}(0) = \hat{v}_{t}/y_t$, and $\hat{\xi}_{t,T}(0) = \hat{u}_{t+1}/y_t$, respectively. The above results applied to MAR(1,0) processes provide a tool for jump detection in the causal autoregressive process with a heavy-tailed error distribution. In each case, the confidence bands are independent of the tail parameter $\alpha$.

$\hat{\xi}_{t,T}(0)$ can be used as a bubble detection tool in the MAR(1,1) and MAR(0,1) processes, as it is constant and close to 0 during the bubble growth and decline periods, and it is time-varying otherwise. Hence, the first difference $\Delta \hat{\xi}_{t,T}$ is also constant and close to 0 during the bubble growth and decline periods.

Consider a set of $\hat{\xi}_{t,T}(0)$, evaluated at $t=1,2,..$  following a high threshold value $y_t$. A close to zero value of $\hat{\xi}_{t+1,T}(0)$ following a large $y_t$ is a warning of an upcoming bubble. Since this statistic remains close to 0 throughout the duration of a bubble, the number of values of $\hat{\xi}_{t,T}(0)$, at $t=t+1, t+2,...$ that are not statistically significant, is a measure of the duration of that bubble. The first time $t_j$ when $\hat{\xi}_{t_j,T}(0) \approx 0$ marks the start of the bubble. The last time $t_J$, such that $\hat{\xi}_{t_J,T}(0) \approx 0$  marks the end of the bubble. 

\subsection{Illustration}

\begin{figure}[H]
     \centering
     \caption{Bubble detection in MAR(1,1) process}
     \begin{subfigure}[b]{0.49\textwidth}
         \centering
         \includegraphics[width=8cm, height = 3cm]{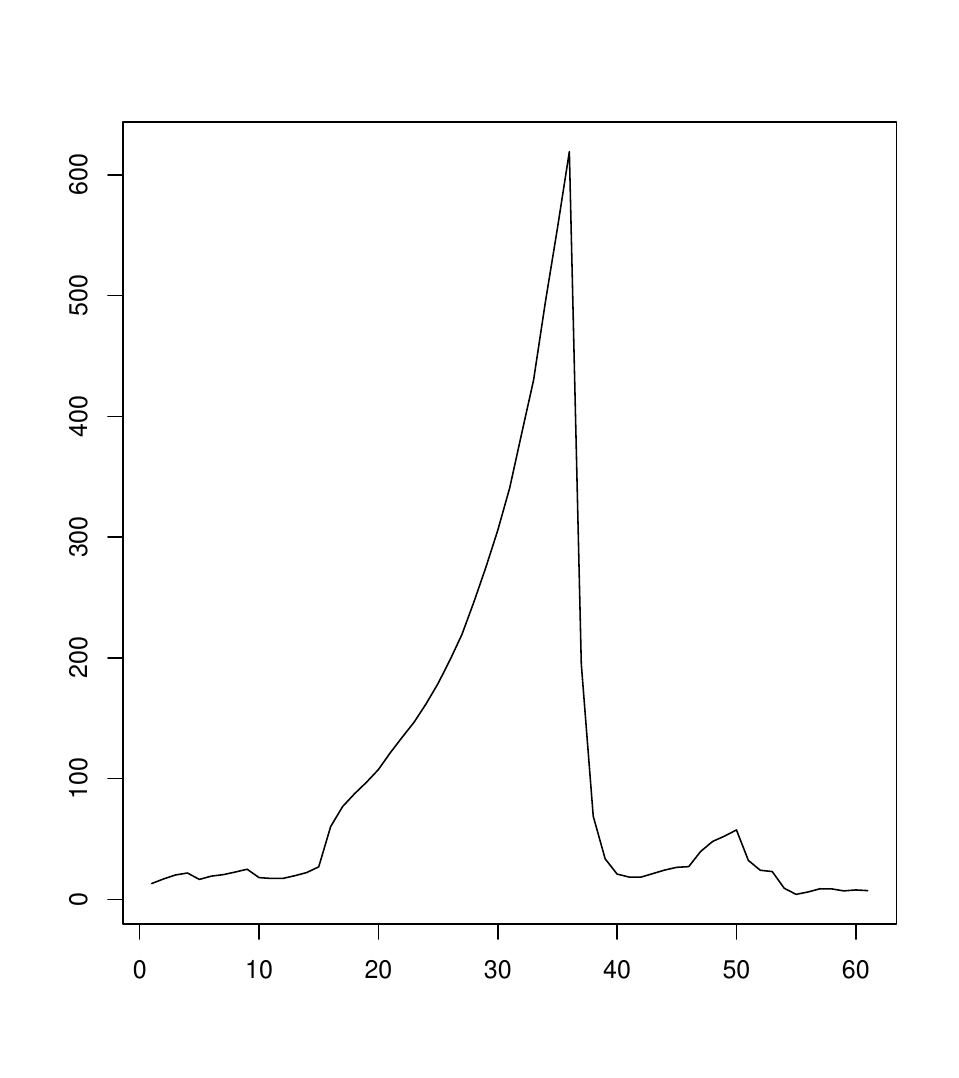}
         \caption{MAR(1,1) process}
     \end{subfigure}
     \hfill
     \begin{subfigure}[b]{\textwidth}
             \centering
             \includegraphics[width=0.49\linewidth, height=3cm]{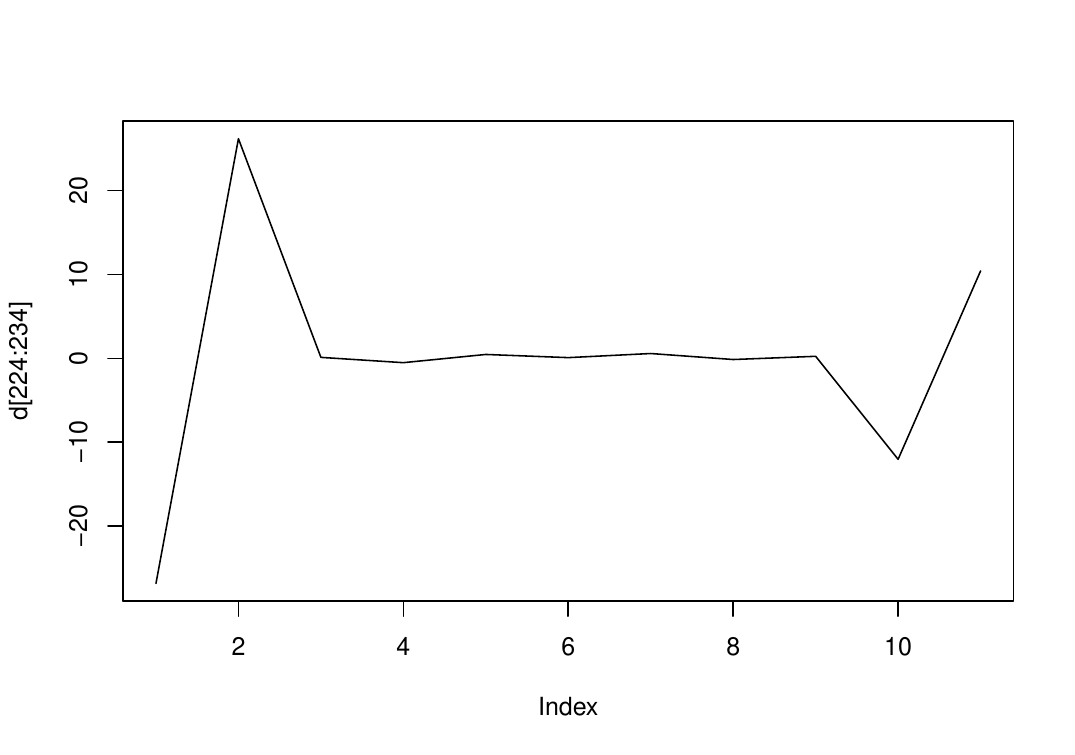}
             \caption{Path of $\Delta \hat{\xi}_{t,T}(0)$}
             \label{fig:enter-label}
     \end{subfigure}   
     \label{fig:bubble}
 \end{figure}

\nin Our approach is illustrated in Figure \ref{fig:bubble} by a bubble episode of a simulated MAR(1,1) process with $\psi=0.9, \phi=0.3$ and Cauchy distributed errors with scale coefficient 1. The top panel of Figure \ref{fig:bubble} displays the bubble episode of the MAR(1,1) process. The first difference $\Delta \hat{\xi}_{t,T}(0)$ of the statistic, defined as:

$$\Delta \hat{\xi}_{t,T}(0) = \Delta \left( \frac{\hat{u}_{t+1} \hat{v}_t} {y_t^2} \right),\; y_t \neq 0$$

\nin is computed and displayed graphically to detect periods when it is approximately constant and close to 0.  The statistic $\Delta \hat{\xi}_{t,T}(0)$ indicates the times when the process becomes a "tail process", which can be used to approximate the start and end of a bubble. The bottom panel of Figure \ref{fig:bubble} shows the first difference $\Delta \hat{\xi}_{t,T}(0)$. The first difference $\Delta \hat{\xi}_{t,T}(0)$ is approximately constant and close to zero during the entire bubble episode.

Figure 3 below shows an example of the path of a simulated MAR(1,1) process with $\phi=0.3$ and $\psi = 0.9$, and i.i.d. errors with a t(3) distribution.

\begin{figure}[h]
        \centering
      \includegraphics[width=13cm, height=5.0cm]{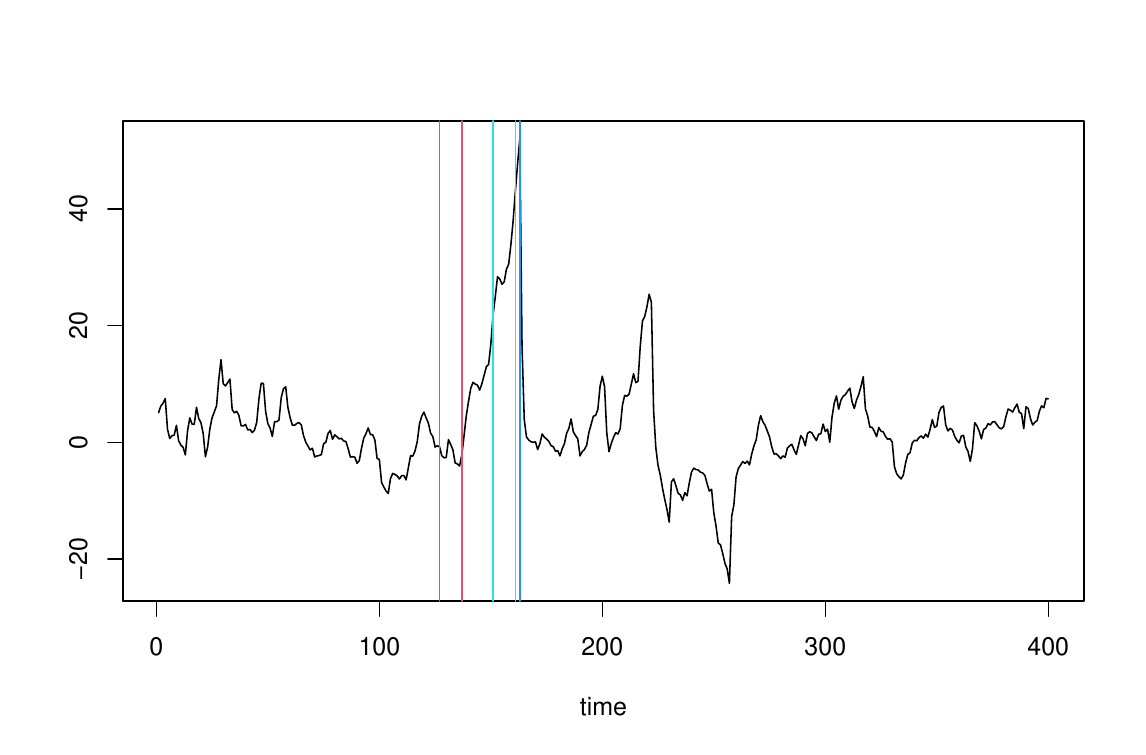}
             \caption{Simulated Path\\
             The vertical lines mark times 127, 137 (red), 151, 161 (blue), 163(green)}      
\end{figure}

We observe a bubble with a peak of 52.051 at $t=163$.
The estimates of the GCov parameter of the MAR(1,1) model are $\hat{\phi}$=0.3085 and $\hat{\psi}=$0.908, with standard errors of 0.028 and 0.030. These estimates are based on $K=4$ power transformations and $H=3$. The test statistic $\hat{\xi}_{t,T}(0)$ evaluated during the bubble, conditional on observation $y(160)=37.951$ is -0.0220 and is within the confidence interval of $\pm 0.05646 $. Therefore, the null hypothesis $\theta_1 = 0$ at time $t=160$ is not rejected.

Next, we examine the effect of the conditioning values and perform functional diagnostics using $\hat{\xi}_{t,T}(h)$ on increasing horizons $h$=1 to 10, conditional on $y_{127}$ =-0.7522  and $y_{151}$ =21.810. The results are reported in Table \ref{tab:simt3}.
The columns of Table \ref{tab:simt3} report: the horizon (col. 1), the value of the test statistic (cols. 2 and 5), the confidence interval of the form $0.0 \pm CI$, and the result of the test of $H_{0,h}: \theta_h=0$ coded 1 for not rejected and 0 for rejected.
We observe that, conditional on the large value at time $t=151$, the diagnostics do not reject the null hypothesis of a bubble over the upward-sloping sequence of the next 10 observations. Conditional on $y_t$ being close to the mean value of the process at time $t=127$, the diagnostics reject the null hypothesis of a bubble over the next 10 observations that are close to the mean.

To evaluate the performance of the test in finite samples, we perform the following experiment. We generate MAR(0,1) processes in samples of length T=400 \footnote{We generate a series of length 800 and discard the first and last 200 observations in 1000 replications}, with t(3), t(4), and t(5) distributed errors and with different values of the noncausal the coefficient $\psi$. We choose samples of 400 observations to increase the probability of occurrence of at least one bubble in a simulated path, which we do not observe. Moreover, we want to increase the probability that the order statistic-based quantile estimator used as a threshold is sufficiently close to the true quantile.

\begin{table}[hbt!]
\centering
\caption{Diagnostics at horizons 1 to 10}
\begin{tabular}{|c|l|c|c|c|c|c|}
\hline
horizon & \multicolumn{3}{|c|}{Conditional on $y_{127}$} & \multicolumn{3}{|c|}{Conditional on $y_{151}$} \\
\hline
h &  $\hat{\xi}_{t,T}(h)$    & CI   &  test  & $\hat{\xi}_{t,T}(h)$ &  CI    & test  \\
\hline
1  & -4.7826 & $\pm$ 0.5263 & 0 & -0.0322 & $\pm$ 0.0581 & 1 \\
2  & -4.5177 & $\pm$ 0.5720 & 0 & -0.0365 & $\pm$ 0.0746 & 1 \\
3  & -4.1350 & $\pm$ 0.5166 & 0 & -0.0340 & $\pm$ 0.0686 & 1 \\
4  & 2.9267 &  $\pm$ 0.1242 & 0 & -0.0326 & $\pm$ 0.0637 & 1 \\
5  & -1.1938 & $\pm$ 0.1028 & 0 & -0.0338 & $\pm$ 0.0671 & 1 \\
6  & -2.7478 & $\pm$ 0.2047 & 0 & -0.0374 & $\pm$ 0.0799 & 1 \\
7  & -7.2927 & $\pm$ 1.1972 & 0 & -0.0377 & $\pm$ 0.0829 & 1 \\
8  & -6.1137 & $\pm$ 1.0591 & 0 & -0.0432 & $\pm$ 0.1054 & 1 \\
9  & -6.7780 & $\pm$ 1.2766 & 0 & -0.0485 & $\pm$ 0.1324 & 1 \\
10 & -2.0080 & $\pm$ 0.2329 & 0 & -0.0552 & $\pm$ 0.1704 & 1 \\
\hline
\end{tabular}
\label{tab:simt3}
\end{table}

\nin We use an upper quantile $y=q_y(0.975)$ as a conditioning large positive threshold, consistently with the statistical literature on exceedances over a high threshold [see e.g. \cite{davis2018inference}]. Next, we compute the statistic at time $t$ from each simulated path, to study the size of the test. To study the power of the test, we use $y=q_y(0.525)$ as the conditioning value at time $t$ followed by a moderate decrease or increase, and preceded by a random pattern, where by moderate we mean that $y_{t+1}$ is less than or equal to $q_y(0.975)$, and $y_t$ is not zero.
The empirical size and power of the test are reported in Tables 4 and 5, Appendix C. 
Table 6 considers the MAR(1,1) with Cauchy distributed errors.
A similar exercise is performed for the MAR(1,1) processes with t-student distributed errors, estimated by the GCov with $K=2$ power transformations and lag $H=4$. The size and power are reported in Tables 7 and 8, Appendix C.

There are some challenges involved in this analysis. It is difficult to accurately estimate the quantiles of the marginal distribution of the process from 400 observations, especially for processes with high persistence and error distributions with low degrees of freedom. Hence, the estimates of $y=q_y(0.975)$ as the conditioning value, obtained from the replicated paths of the process, may vary. Moreover, without observing the trajectory, we do not know a priori
whether the process admits a bubble or a jump instead, the latter occurring in the MAR(1,1) processes due to the causal component. 
In processes with $\psi=0.9$, bubbles are infrequent and grow slowly at a rate of about 1.1. 
In processes with $\psi=0.6$, the bubbles are more frequent and grow faster. We distinguish the bubble from jumps by checking if, following the conditioning value of $y=q_y(0.975)$, the process grows at a rate close to the theoretical rate.
Another difficulty is the asymptotic normality of the estimators, which impacts the results in the MAR(0,1) processes with t(3) distributed errors estimated by the OLS. 

In Table 4, we observe that the test over-rejects when the error distribution is t(5), especially for higher values of $\psi$. Table 5 shows that the power of the test is slightly lower when the noncausal persistence is higher. Table 6, illustrating processes with $\psi=0.9$ and Cauchy distributed error, shows that the test tends to under-reject for closer to 0.9 values of parameters $\phi$. In addition, the power of the test decreases for higher values of $\phi$.  
In Table 7, we observe that the size of the test is closer to the nominal value for moderate values of $\psi = 0.7$ and $\psi=0.8$. The test is conservative for $\psi=0.6$
and over-rejects for $\psi=0.9$ in processes with t(3) distributed errors. For each $\psi$, the test is the most conservative in processes with t(5) distributed errors. This error distribution is the closest to the normal among the three error distributions considered, leading to potential parameter identification problems. We find that the size is close to the nominal value for moderate values of causal persistence $\phi$, regardless of $\psi$. Table 8 shows that the test has
good power, especially for $\psi = 0.7$ and $\psi=0.8$.  The power deteriorates in processes with t(5) distributed errors for $\psi=0.6$ and $\psi=0.9$ and for higher values of causal persistence, making bubbles harder to distinguish from standard dynamics.

\section{Green Stock Indexes and ETFs}
\label{sec:GreenUni}

The transition to a clean energy economy has recently gained significant attention, driven by global events such as the COVID-19 pandemic and the Russia-Ukraine war. %The COVID-19 pandemic, which began in late 2019, caused widespread disruptions in global economies and supply chains, including those in the energy sector. The pandemic exposed the fragility of traditional energy systems, particularly those reliant on fossil fuels. Lockdowns and restrictions led to a dramatic reduction in energy demand, highlighting the volatility of oil prices and the dependence on non-renewable energy sources; see, among others, \cite{bourghelle2021oil}, \cite{ghabri2021fossil}. Similarly, the Russia-Ukraine war, which intensified in 2022, has significantly impacted global energy markets. As a major supplier of oil and natural gas, Russia's actions have disrupted energy supplies, causing price spikes and shortages in countries dependent on Russian energy exports [\cite{chen2023russia}]. 
In fact, in response to these crises, governments, industries, and investors have increasingly prioritized clean energy investments, recognizing their long-term benefits for a stable and environmentally friendly energy system [\cite{mohammed2023all}]. This growing focus on clean energy is also driven by the need to achieve net zero emissions by 2050, which requires substantial investment in clean energy from both developed and developing countries [\cite{khalifa2022accelerating}]. In 2023, global investment in the energy sector was estimated at USD 2.8 trillion, an increase of 0.6 trillion USD from five years earlier. Almost all of this growth was directed toward clean energy and infrastructure, increasing total clean energy spending to 1.8 trillion USD, compared to 1 trillion USD for fossil fuels\footnote{IEA (2023), World Energy Outlook 2023, IEA, Paris https://www.iea.org/reports/world-energy-outlook-2023, License: CC BY 4.0 (report); CC BY NC SA 4.0 (Annex A).}. These large investments pose the risk of "green bubbles" -- rapid stock price increases followed by crashes. Specifically, such bubbles occur when overinvestment and speculative behavior drive the market value of clean energy assets beyond sustainable levels. Consequently, rising interest rates can increase financial pressure on investors, potentially triggering bubble bursts and undermining the credibility of the clean energy transition [\cite{wimmer2016green}].

The literature on green energy stocks is relatively recent and focuses on return analysis. The pioneering paper by \cite{henriques2008oil} shows that returns on technology stocks and oil prices are both Granger-caused by the returns of green (alternative) energy stocks, based on a vector autoregressive (VAR) model. The relationship between green stocks and oil has also been examined by \cite{sadorsky2012correlations}, \cite{sadorsky2012modeling}, and  \cite{kumar2012stock}. These articles consider carbon prices and do not find their significant impact on green energy stocks. In contrast, this impact is evidenced after the year 2007 by \cite{managi2013does}.  
The literature has not yet examined the green stock price dynamics to detect and explain the presence of bubbles, which is done in this paper.

\subsection{The Data}

We consider the Renixx Index, the WHETF, and the iShare green stock ETFs. For Renixx, we use daily data from \url{https://www.renewable-energy-industry.com/stocks} and sample them at a monthly frequency by taking the last day of each month of the closing price series. When the last day of the month is a bank holiday, we consider the previous available day. The Renixx index tracks the global renewable energy market, covering sectors such as wind, solar, bioenergy, geothermal, hydropower, electronic mobility, and fuel cells. It comprises 30 companies, each of which derives more than 50\% of its revenues from these sectors (see \url{www.iwr.de/renixx}). The WHETF tracks the WilderHill Clean Energy ETF, which includes companies listed in the United States that focus on developing cleaner energy and conservation efforts. In particular, WHETF allocates a minimum of 90$\%$ of its total assets to common stocks included in this ETF. It is rebalanced and reconstituted quarterly. The iShare tracks the S$\&$P Global Clean Energy ETF, which provides exposure to the top 30 largest and most liquid publicly traded companies operating worldwide in the clean energy sector, based on a modified market capitalization weighting system.

We investigate Renixx on its whole available sample from January 2002 to February 2024, and both WHETF and iShare from January 2009 to February 2024. 
%to avoid the latter two ETFs starting the sample in the middle of the financial bubble. 
We consequently have a total of $T=266$ monthly observations for Renixx and $T=182$ observations for WHETF and iShare, with potentially two bubble patterns for Renixx and a single bubble for both the WHETF and iShare. 
The Renixx index experienced significant bubbles in 2008 and 2020, coinciding with two major global events: the 2008 financial crisis and the outbreak of the COVID-19 pandemic. In 2008, the financial crisis rocked global markets, leading to widespread economic instability and investor panic. The collapse of major financial institutions, coupled with a credit crunch and falling stock markets, triggered a flight to safety among investors. This risk aversion had a significant impact on the renewable energy sector, with reduced investment in renewable energy projects and a decrease in the demand for renewable energy stocks [see \cite{giorgis2024salvation}]. Similarly, in 2020, the COVID-19 pandemic caused unprecedented economic disruption around the world. Lockdown measures, supply chain disruptions, and reduced consumer spending resulted in a global economic downturn. The renewable energy sector was particularly affected by the decrease in energy demand due to reduced economic activity and travel restrictions. 
Figure \ref{fig:renixx1} shows the path of  Renixx between January 2009 to February 2024. WETHF and iShare are displayed in Figures \ref{fig:WHETFDet} and 4c, respectively. However, only the COVID period bubble appears in the data, as the observations are from January 2009 to February 2024, and therefore do not include data from the global financial crisis. 

\subsection{Prices vs. Returns}
\begin{figure}[H]
    \centering
    \caption{Green Index and ETFs from January 2009 to February 2024. The panels on the left show the original series, while those on the right display their detrended versions.}

    \label{fig:green}
    \begin{subfigure}[b]{\textwidth}
        \centering
        \includegraphics[width=13cm, height=3.3cm]{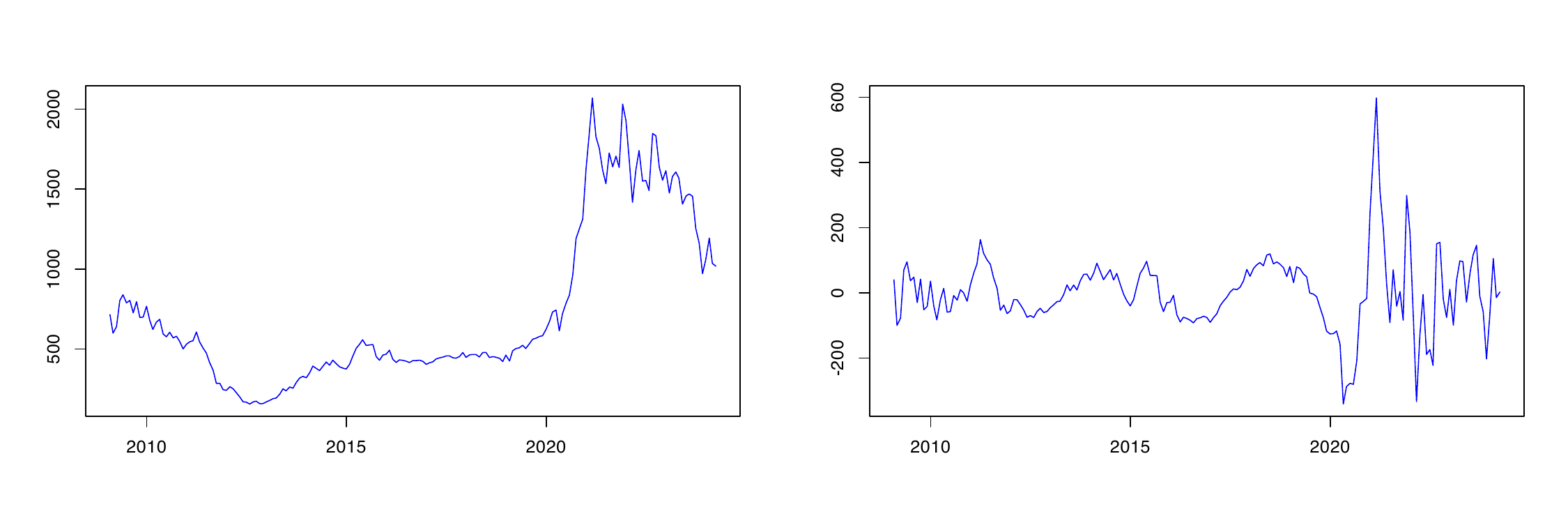}
        \caption{Renixx}
        \label{fig:renixx1}
    \end{subfigure}
    
    \vspace{-0.3cm} % spazio verticale tra i grafici
    
    \begin{subfigure}[b]{\textwidth}
        \centering
        \includegraphics[width=13cm, height=3.3cm]{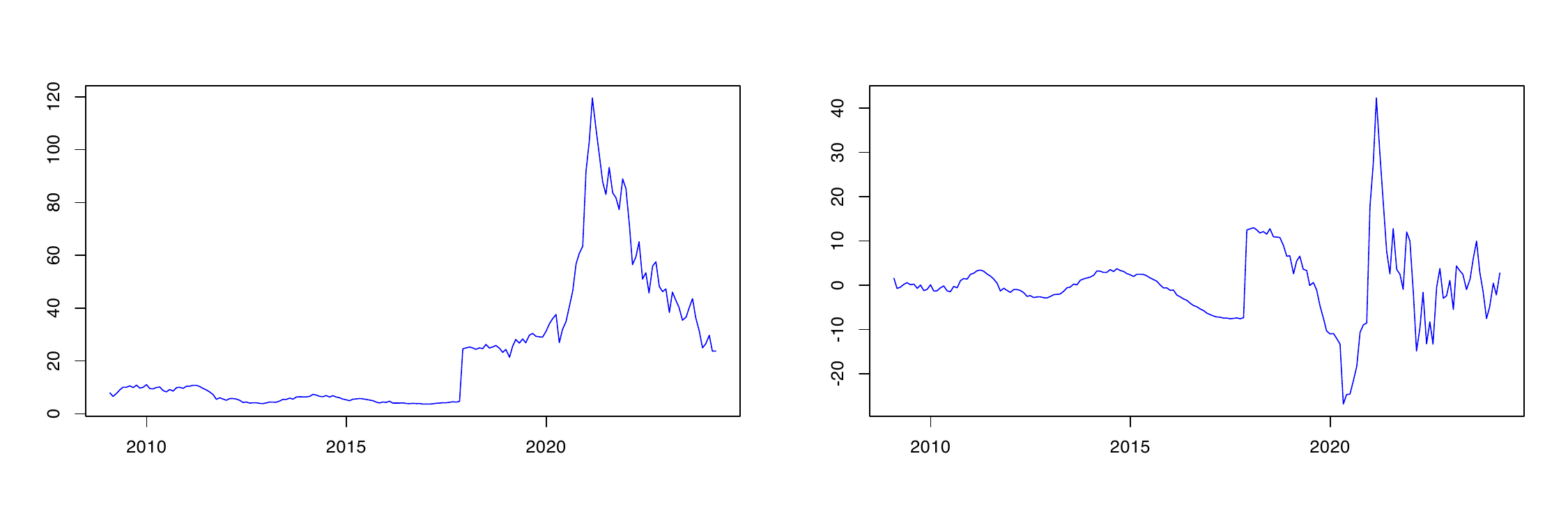}
        \caption{WHETF}        
        \label{fig:WHETFDet}
    \end{subfigure}
    
    \vspace{-0.3cm}
    
    \begin{subfigure}[b]{\textwidth}
        \centering
        \includegraphics[width=13cm, height=3.3cm]{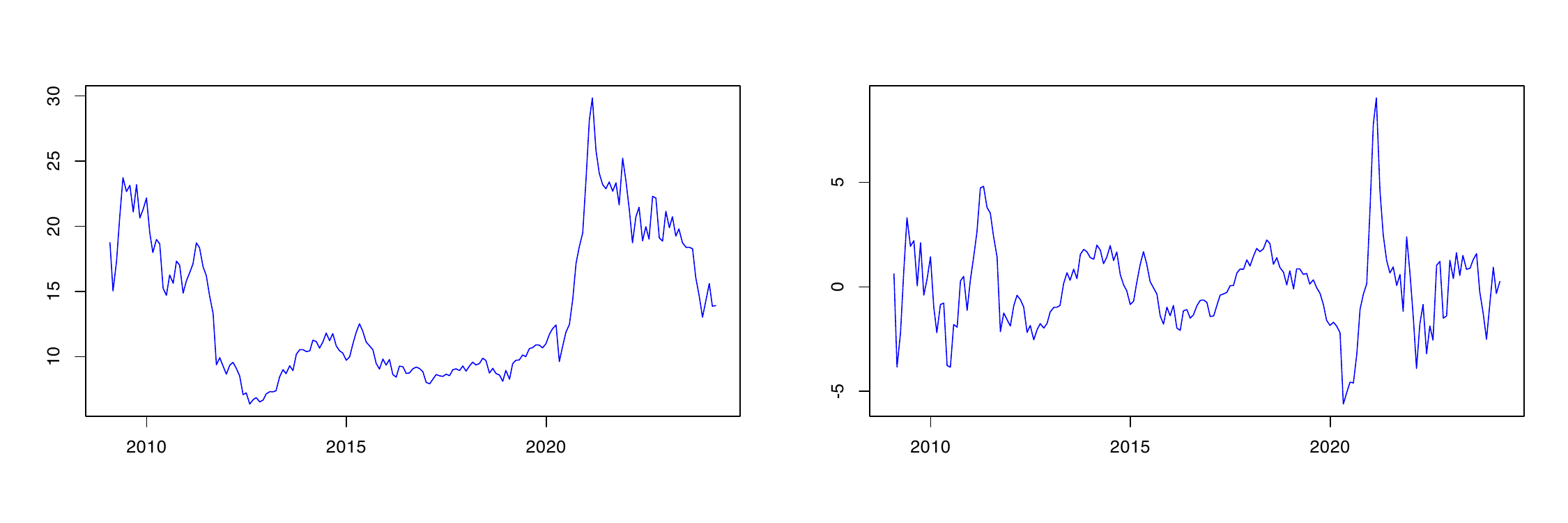}
        \caption{iShare}        
        \label{fig:WHETFDet2}
    \end{subfigure}
\end{figure}

 Although most studies in green energy finance focus on returns [e.g., \cite{henriques2008oil}; \cite{sadorsky2012correlations}], our analysis is based on prices, allowing for a direct examination of bubble dynamics that would otherwise be obscured by log-differencing. In fact, in financial analysis, especially in commodity pricing, it is crucial to focus on the price process rather than returns or first differences. This approach aligns with the financial theory that underlies commodity pricing, as discussed in \cite{hull2016options}. Transforming the price series into returns or first differences can obscure critical aspects of the price process, including the presence of bubbles. Moreover, differencing can eliminate noncausal components essential to understanding bubble dynamics [\cite{giancaterini2022climate}]. We employ the detrended cubic spline method to address this issue and eliminate trend components while preserving significant bubble patterns, as detailed in \cite{jasiakhall}. This method fits a cubic spline, a piecewise function composed of polynomial segments of degree three, to the time series data. The points where these segments connect, known as knots, allow separate cubic polynomials to be fitted within each segment. We place knots every two years to effectively detrend the series, balancing data smoothing and avoiding overfitting. By applying this approach, we successfully isolate the cyclical variations that contain the bubble patterns from the trend component. The detrended series are shown in the right panels of Figure \ref{fig:green}.\footnote{An alternative detrending technique that may help us preserve the bubble patterns is the HP filter method [see \cite{giancaterini2022climate}, \cite{hecq2023predicting}]. However, as indicated by \cite{jasiakhall}, the HP filter requires you to choose a value for the smooth parameter lambda, which is typically a function that increases with the sampling frequency of the data. In our examination of monthly data, a high lambda value may result in computational inaccuracies, thus justifying our preference for using a cubic spline to detrend the data. However, we have investigated several detrending methods, including the HP filter, as well as polynomial trends of different degrees (results upon request). We only report the results with the spline detrending approach as it passes the absence of nonlinear serial dependence test [\cite{jasiak2023gcov}]. Finally, note that an alternative detrending approach based on unobserved components has been developed by \cite{Blasques2023Observation}}.

\subsection{Summary Statistics and MAR Estimations}

 The detrended data is non-Gaussian, as evidenced by the Kolmogorov-Smirnov and Shapiro-Wilk's tests, which both reject the null hypothesis of normality.
 Table \ref{tab:stats} presents the summary statistics, which confirm that the distributions of the series are non-Gaussian, given the reported excess kurtosis. We test the spline-detrended data for causal and noncausal persistence by using the test introduced in \cite{jasiak2023gcov} [see Section \ref{sec:GCov}]. We choose $K=2$, including the time series and its squares as (non)linear transformations, and $H=3$ as the number of lags in the objective function of the test. The null hypothesis of the absence of nonlinear serial dependence is rejected since the test value is 514.98, while the critical value at a 5$\%$ significance level from the chi-square distribution is $\chi^2(12)$ = 21.026.

\begin{table}[hbt!]
\centering
\caption{Summary statistics of Renixx, WHETF, and iShare}
\begin{tabular}{cccccccc}
\toprule 
       & $T$ & Mean    & Min      & Max     & SD     & Skewness & Ex.Kurt. \\ \midrule
Renixx & 266   & 748.57	&155.47	&2070.38	&475.09	&1.03 &	2.85  \\ 
WHETF  & 182   & 23.08&	3.66&	119.57	&25.31	&1.67	&5.28    \\ 
iShare & 182   & 13.60 &	6.36&	29.83&	5.46&	0.74	& 2.36   \\ \bottomrule
\hline
\end{tabular}
\label{tab:stats}
\end{table}

We apply the GCov estimator to obtain the parameter estimates of MAR models for the three series. Specifically, we use $H=K=2$ and 
$a_j(\epsilon_t)=\epsilon_t^j$, for $j=1,2$, in \eqref{eq:GCov22}, i.e., the residuals and their squares as transformations in the GCov estimation of Renixx [see \cite{cubadda2023optimization}] 
and 
$a_j(\epsilon_t) = \log(|\epsilon|)^j$ for $j = 1, 2$, for WHETF and iShare.
%as this logarithmic transformation helps mitigate the computational problems associated with the variances. 
Table \ref{tab:UniEst} presents the estimation results. To identify the dynamics of the underlying processes, we evaluate all combinations of $r$ and $s$ such that $r+s=p$. We begin with $p=1$ and gradually increase $p$ to find the values of $r$ and $s$ that provide i.i.d. residuals based on the GCov specification test. The last row of  \ref{tab:UniEst} shows the results of the GCov test. This approach allows us to identify the Renixx index and the two ETFs as MAR($1,1$) processes.   The GCov specification test results indicate that these processes are correctly specified and provide a good fit to the data.

\begin{figure}[H]
\caption{Causal and noncausal components}
    \centering
    % First row with two subfigures
    \begin{subfigure}[t]{0.33\textwidth}
        \centering
        \includegraphics[width=\textwidth, height=4 cm]{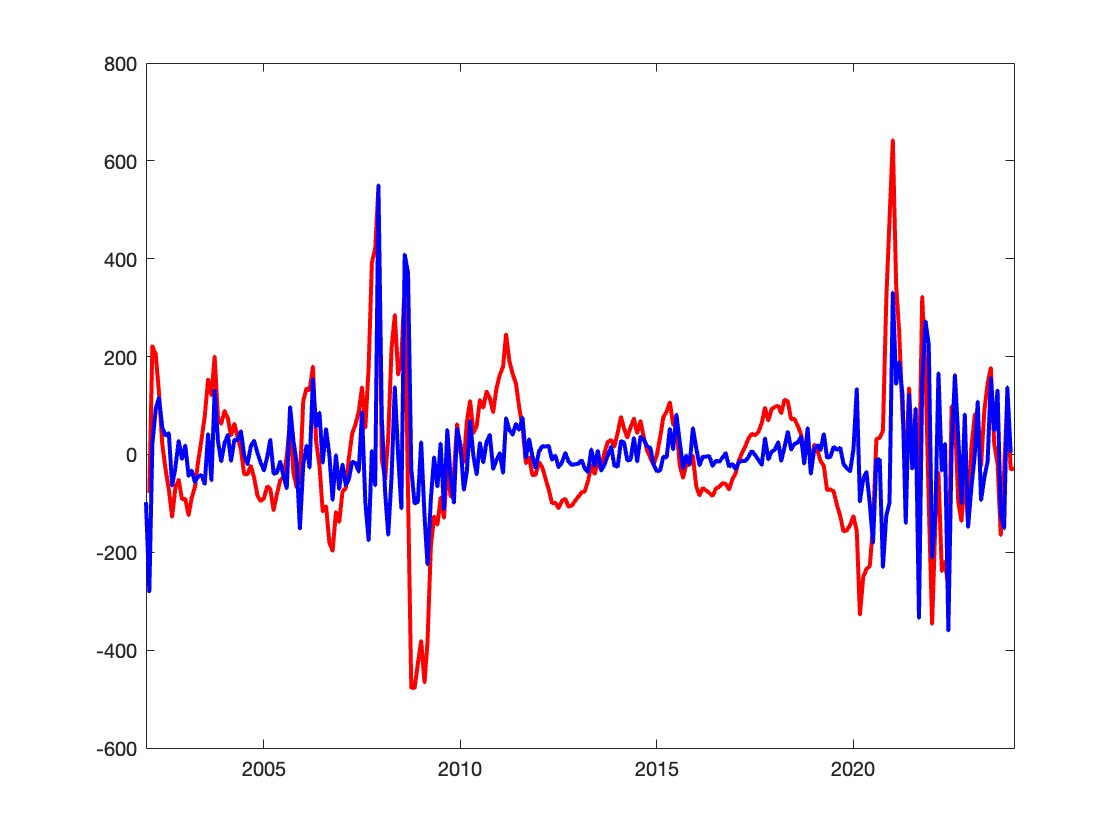}
        \caption{Causal (blue) and Noncausal (red) components of Renixx}
        \label{fig:renixx}
    \end{subfigure}%
    \hfill
    \begin{subfigure}[t]{0.33\textwidth}
        \centering
        \includegraphics[width=\textwidth, height=4 cm]{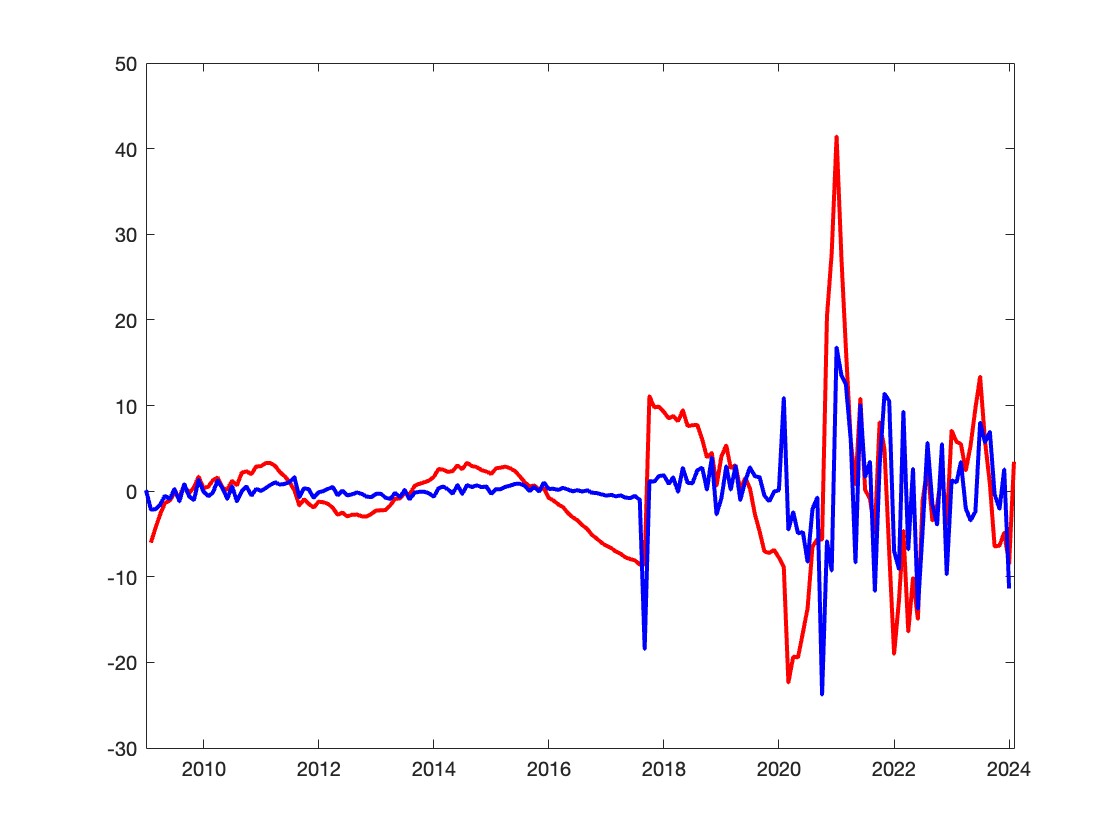}
        \caption{Causal (blue) and Noncausal (red) components of WHETF}
        \label{fig:wthef}
    \end{subfigure}
    \hfill
    % Second row with one centered subfigure
    %\vspace{0.5cm} % Space between rows
    \begin{subfigure}[t]{0.33\textwidth}
        \centering
        \includegraphics[width=\textwidth, height=4 cm]{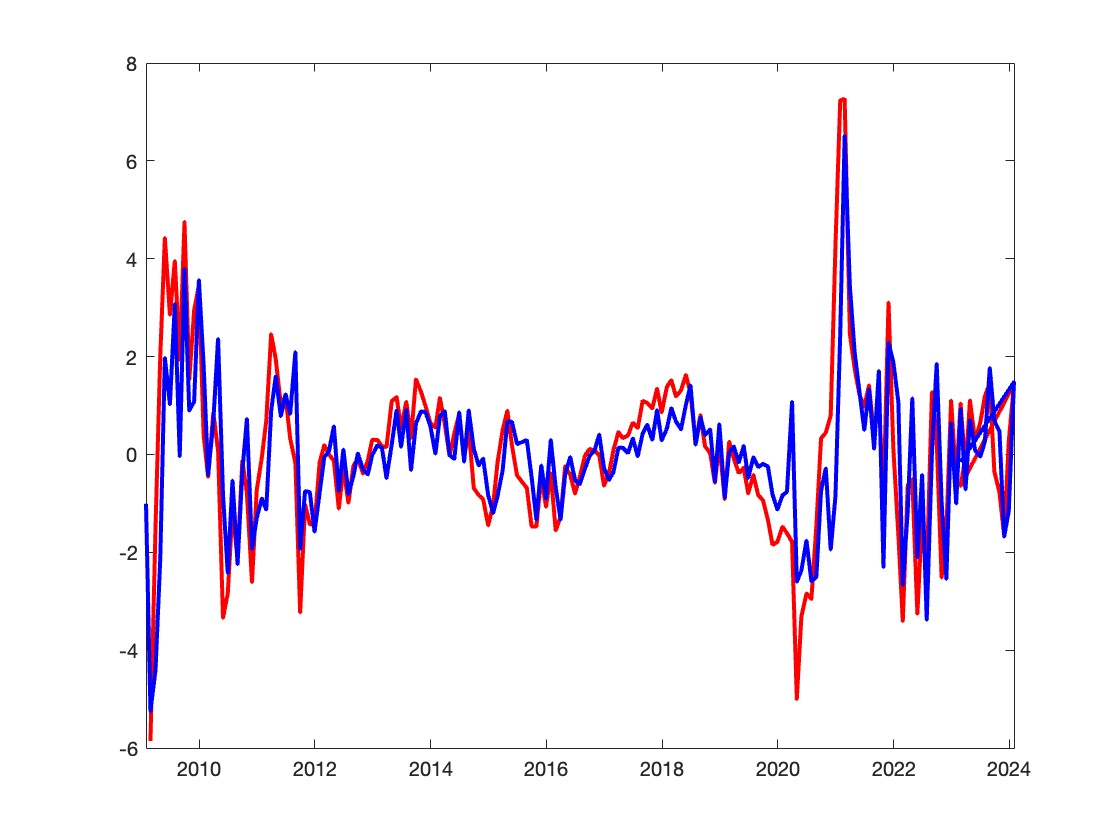}
        \caption{Causal (blue) and Noncausal (red) components of iShare}
        \label{fig:ishare}
    \end{subfigure}
    \label{fig:combined}
\end{figure}

Figure 5 shows the estimated latent components $\hat{u}_t$ and $\hat{v}_t$ defined in eq. (3) and (4) for the three series of interest. We observe that the changes in the noncausal component anticipate the dynamics of the observed series, followed by the causal component. In the next section, these estimates are used to compute the statistic $\hat{\xi}_{t,T}(0)$ for each series.

\begin{table}[H]
\caption{\textit{Estimated MAR(1,1) coefficients and GCov tests}}
\centering
\resizebox{0.6\textwidth}{!}{
\begin{tabular}{cccc}
\toprule
MAR($1,1$)     & Renixx & WHETF & iShare \\ 
\midrule
$\phi$                   & 0.24         & 0.07           & 0.32            \\
                         & (0.11)            & (0.01)                & (0.02)      \\
$\psi$                & 0.70           & 0.89           & 0.62            \\
                         & (0.08)            &  (0.03)               & (0.02)      \\
\midrule
GCov Test    & 7.14         & 4.86         & 2.40           \\
(Critical Value)  & (12.59)      & (12.59)      & (12.59)       \\
\bottomrule
\end{tabular}}
\label{tab:UniEst}
\caption*{Top panel: estimates with standard errors in parentheses;
Lower panel: residual-based specification test}
\end{table}

\subsection{Bubble Detection}

\begin{figure}[H]
    \caption{Local variance of time series with $H=5$ rolling window size}  
    \centering
    \begin{subfigure}[b]{0.32\textwidth}
         \centering
         \includegraphics[width=\textwidth, height = 4cm]{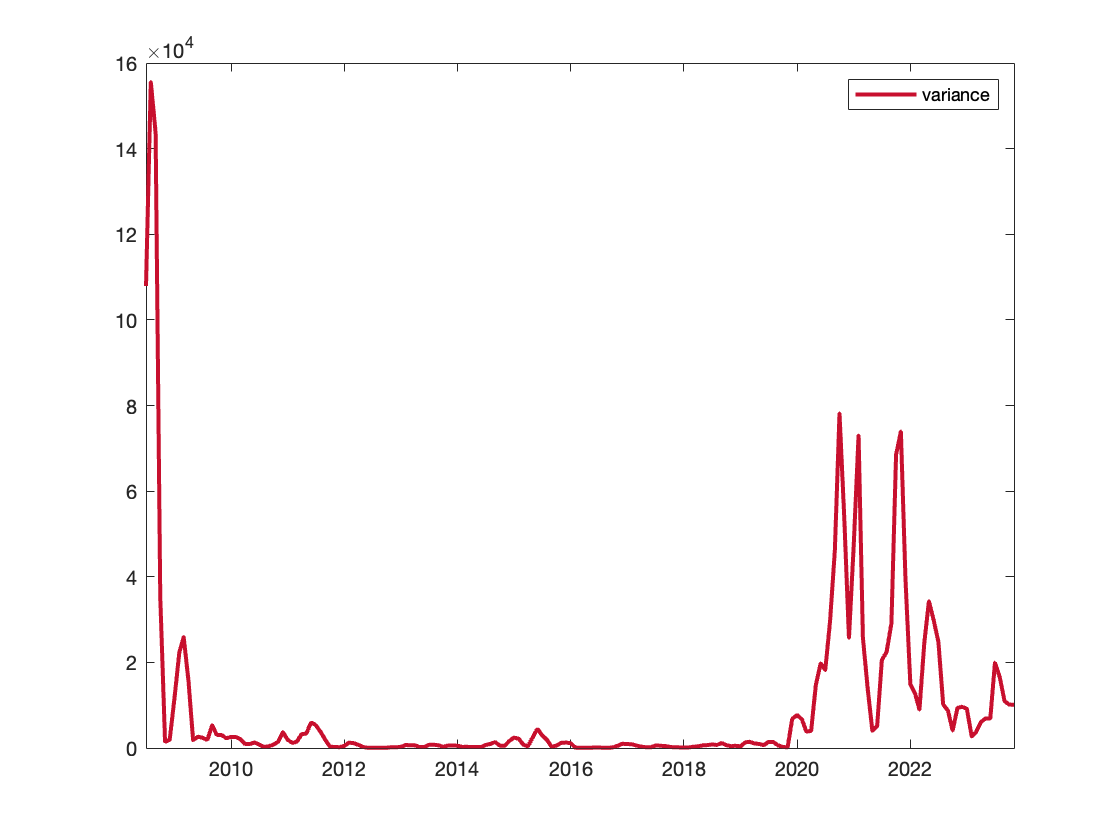}
         \caption{Renixx}
         \label{mean5overlap}
     \end{subfigure}
     \hfill
        \begin{subfigure}[b]{0.32\textwidth}
        \centering
         \includegraphics[width=\textwidth, height = 4cm]{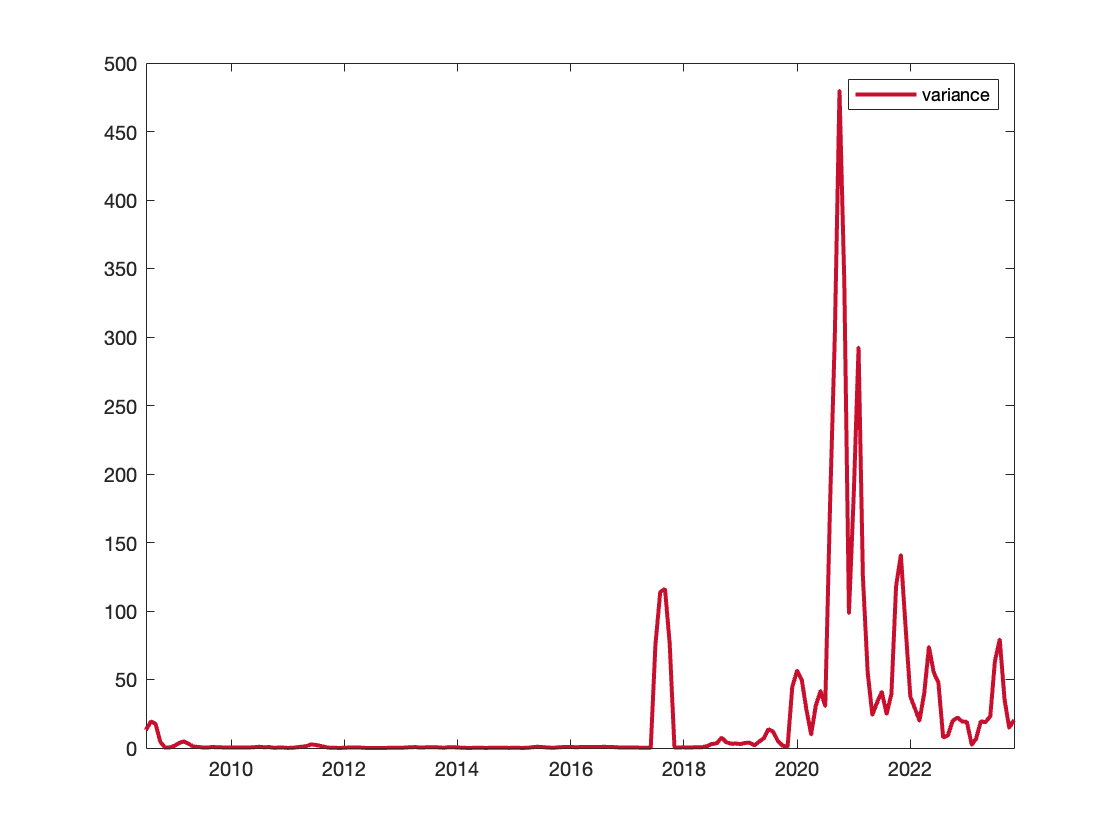}
         \caption{WHETF}
         \label{Cycle_t9}
     \end{subfigure}
     \hfill
     \begin{subfigure}[b]{0.32\textwidth}
        \centering
         \includegraphics[width=\textwidth, height = 4cm]{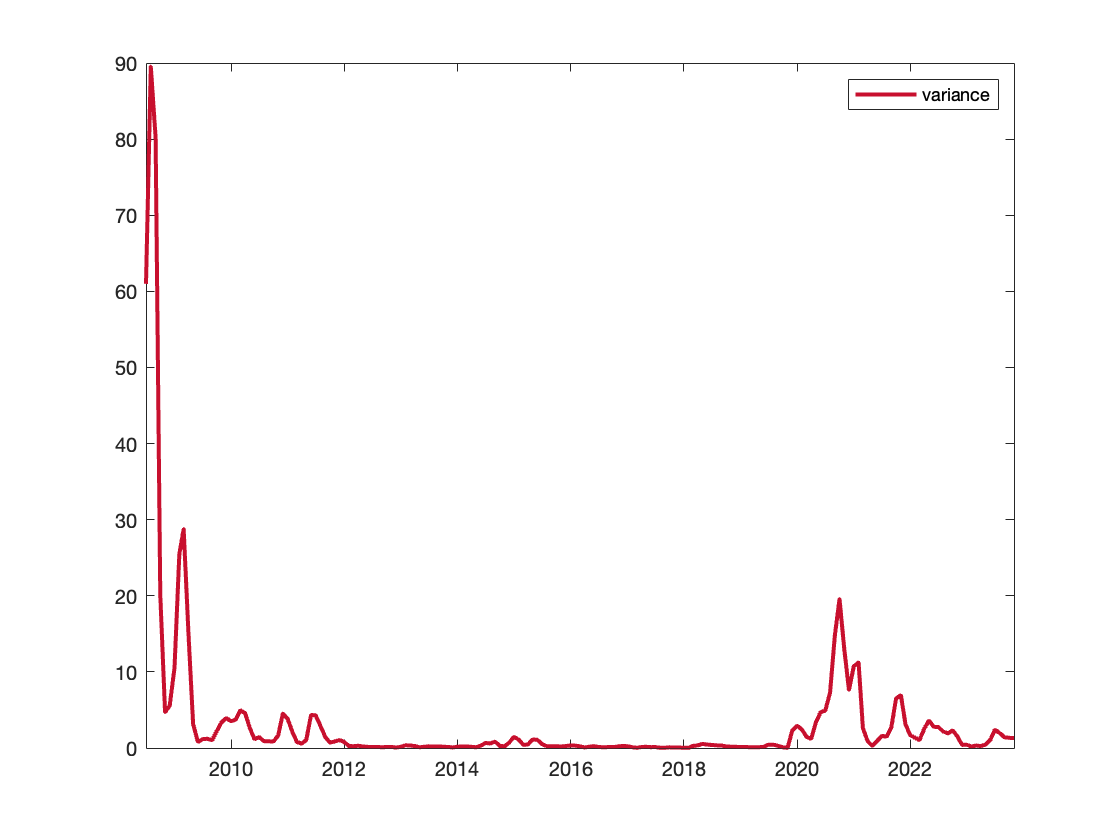}
         \caption{iShare}
         \label{Eigenvalues}
     \end{subfigure}  
     \label{fig:local}
 \end{figure}

The bubbles in the price series can manifest themselves as periods of high volatility. 
We estimate local variances to detect periods of high volatility in Renixx and the ETFs. We consider a rolling window of 5 observations and estimate the local variance for the three series. The plot of rolling estimates provides a graphical tool for preliminary analysis. Figure \ref{fig:local} displays the rolling estimates of the variances of the Renixx index and two ETFs. Two bubble periods, one that occurred during the financial crisis of 2008 and one associated with the COVID pandemic, are observable in Figure \ref{fig:local}. Since the data on ETFs before 2009 are not available and the observations on bubbles during that period are incomplete, we focus on the bubble observed during the COVID period.

\begin{figure}[H]
    \caption{Bubble detection statistics for Renixx, WHETF, and iShare }  
    \centering
    \begin{subfigure}[b]{0.32\textwidth}
         \centering
         \includegraphics[width=\textwidth, height = 4cm]{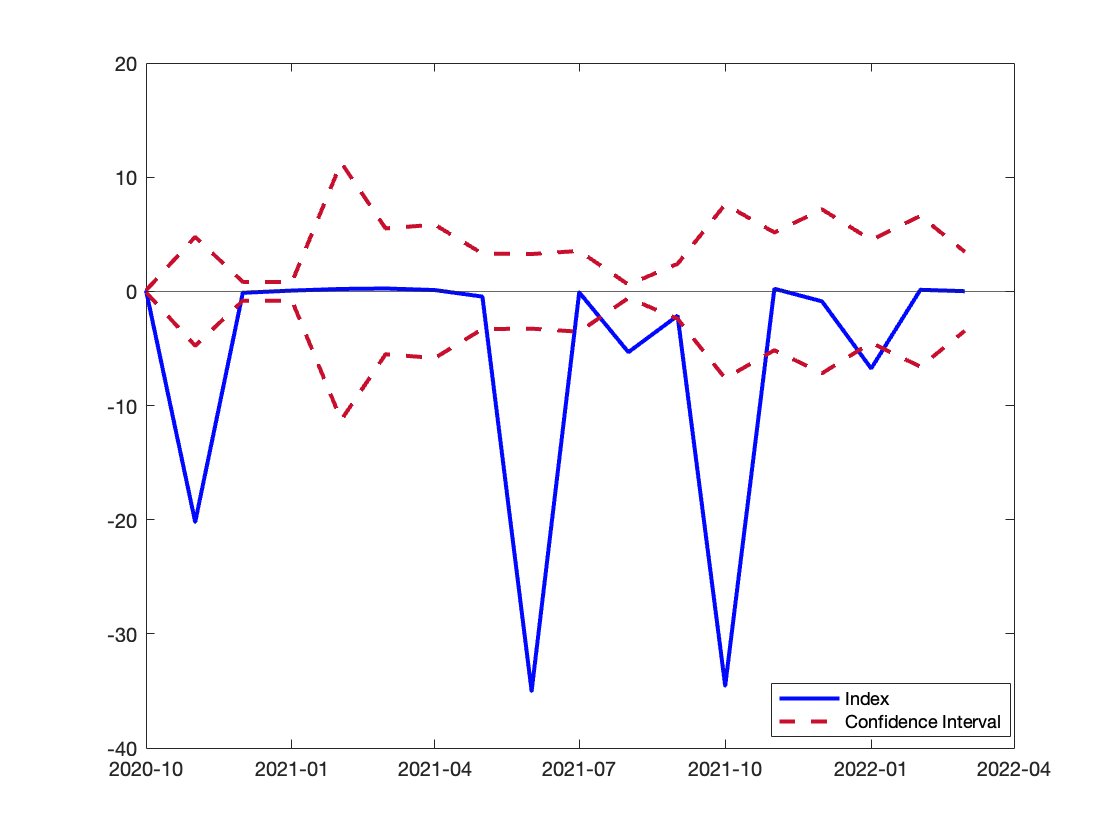}
         \caption{Renixx}
         \label{s-r}
     \end{subfigure}
     \hfill
        \begin{subfigure}[b]{0.32\textwidth}
        \centering
         \includegraphics[width=\textwidth, height = 4cm]{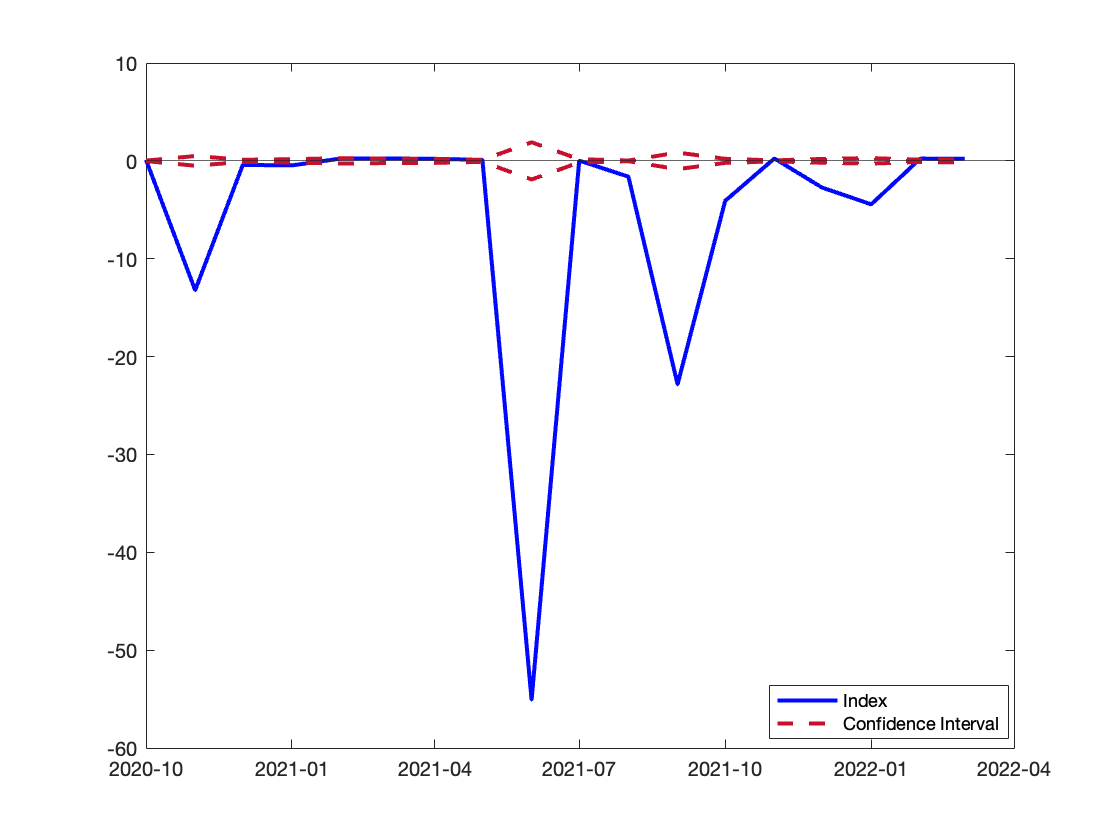}
         \caption{WHETF}
         \label{s-w}
     \end{subfigure}
     \hfill
     \begin{subfigure}[b]{0.32\textwidth}
        \centering
         \includegraphics[width=\textwidth, height = 4cm]{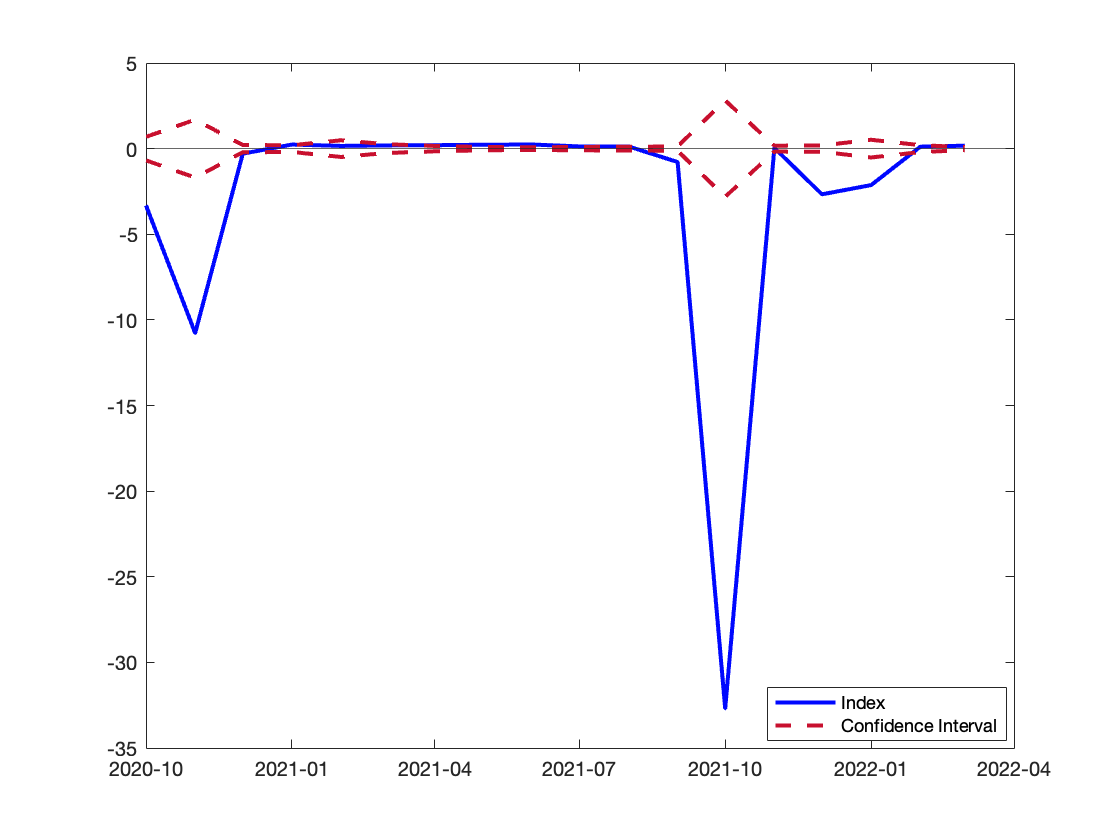}
         \caption{iShare}
         \label{s-i}
     \end{subfigure}  
     \label{fig:newstats}
 \end{figure}

To detect and determine the dates of bubbles in the three price series, we focus specifically on the period between 12/01/2020 and 05/31/2021. Figure \ref{fig:newstats}  illustrates how the method introduced in Section 4 can be applied to Renixx and the ETFs to test for bubbles, and to determine the dates at which that bubble starts and ends. Figures \ref{s-r} and \ref{s-w} show the test statistics $\hat{\xi}_{t,T}(0)$ computed over that period and the associated confidence band.

The conditioning threshold values $q$ used in the analysis correspond to the upper tail quantiles at levels $97.5\%$ in Renixx, $98\%$  in WHETF, and $96\%$ in iShare. We use the nonparametric forward estimator of the conditional distribution function introduced in \cite{davis2018inference} to estimate the probability that $y_{t+1}/y_t \geq 1$  conditional on $y_t>q$, where $q$ denotes the tail quantile from each sample given above. This provides us with the probabilities of tail process dynamics for each time series, following a threshold value $q$. For Renixx and WHETF, the estimated conditional probability is $80\%$, and for iShare it is $40\%$. Given these non-zero conditional probability estimates, we proceed with the analysis\footnote{Note that these results need to be considered with caution as the approach is valid in large samples.}.

We observe that the bubble in Renixx and WTETF started on 12/01/2020 and ended on 03/01/2021. The bubble in iShare ended on 04/30/2021. Note that in Figure \ref{s-i}  the confidence intervals depend on the parameter variances reported in Table \ref{tab:UniEst} and are larger for Renixx compared to the other estimated processes. 

Given a tail index value, the time to peak depends on the rate of bubble growth. It takes longer for the bubble to grow
when $\hat{\psi}$ is large. As shown in Section 3.3, the time to peak is also determined by the tail parameter, which is estimated from the Hill estimator R-package. For $\hat{\alpha}=1.3$, the expected time to peak for Renixx is 1.5 months (computed as the conditional expectation of $N$ given $N<0$), and the probability that $N$ exceeds 3 months is 0.232. 

\noindent Given $\hat{\alpha}=1.7$ for WHETF, the expected time to peak is 4.5 months, and the probability that it exceeds 5 months is 0.37.
With $\hat{\alpha}=1.15$, the time to peak for iShare is slightly longer than 1 month, and the probability that it exceeds 3 months is 0.17.

\section{Conclusions}
\label{sec:Conclusions}

This paper introduced a new statistical test for detecting financial bubbles, derived from the tail process representation of mixed causal-noncausal autoregressive models. Unlike traditional unit root-based methods, our approach is grounded in strictly stationary noncausal dynamics, which provide a natural framework for capturing locally explosive behavior. Classical tests such as SADF or GSADF are often sensitive to heavy tails or may misinterpret short-lived spikes as bubbles. In contrast, our test exploits the tail process properties of causal and noncausal dynamics, and has a statistic remaining approximately constant and close to zero during a bubble episode. The additional inference on time-to-peak and bubble duration allows us to separate persistent speculative bubbles from short-lived spikes and regular fluctuations.  

We applied univariate MAR($r,s$) models to the Renixx index and two green energy ETFs (WHETF and iShare), estimating the underlying dynamics using the GCov approach. This combination provides a robust identification strategy: the GCov estimator ensures reliable parameter estimation under non-Gaussianity, while the proposed test captures the tail process behavior that signals bubble formation. The empirical results reveal bubbles in all three series, showing that the test performs well in detecting and dating speculative episodes in green financial markets.  

For bubble detecting in real time, the proposed method can be applied sequentially when new observations become available. In that case, the model needs to be re-estimated, and the test statistic recalculated at each step.  Then, one may need to adjust the test level to the nominal size using, e.g., the approaches given in \cite{virani2019sequential}. Moreover, the causal-noncausal process can be forecasted out-of-sample [\cite{lanne2012optimal},
\cite{gourieroux2016filtering, gourieroux2026nonlinear}, \cite{hecq2021forecasting, hecq2023predicting}] and the method proposed in this paper can be applied to the time series augmented by the forecast period.

\newpage
\appendix

\newpage
\section*{Appendix A}
\label{App:C}

\nin \textbf{0. AR(p) Representation of the MAR(r,s) Process}

\nin  An alternative representation of the univariate causal-noncausal process (1) is the following AR(p) with $p=r+s$:
\begin{equation*}
    y_t = \varphi_1 y_{t-1} - \dots - \varphi_p y_{t-p} + e_t,
    \label{eq:rootsIO}
\end{equation*}

\nin where $e_t = -\frac{1}{\psi_s} \epsilon_t$ [\cite{brockwell1987stationary}]. In this representation, the polynomial $\varPhi(L) = 1-\varphi_1 L - \cdots - \varphi_p L^p$ has roots both inside and outside the unit circle. Specifically, the roots outside the unit circle correspond to the causal component of $y_t$, while the roots inside the unit circle indicate the noncausal component that captures bubbles and other nonlinear features. Note that in either representation of the model, the error $e_t$ is not an innovation process because $e_t$ is not independent of $y_{t-1}, y_{t-2},...$. 

\color{black}
\medskip

\nin \textbf{1. Derivation of the MA($\infty$) representation of MAR(1,1)}

\nin We have:
$$ (1-\phi L )(1-\psi L^{-1}) = -\psi L^{-1} (1-\phi L)(1-\psi^{-1}L).$$

\nin It follows using the partial fraction decomposition of the polynomial that:
\begin{eqnarray*}
\Frac{1}{(1-\phi L )(1-\psi L^{-1})}  & = & -\psi^{-1} L \Frac{1}{(1-\phi L )(1-\psi^{-1} L)}\\
& = & \Frac{- \psi^{-1} L}{\phi - \psi^{-1}}
\left( \Frac{\phi }{1-\phi L} - \Frac{\psi^{-1}}{1-\psi^{-1} L} \right) \\
& = & \Frac{1}{1 - \phi \psi} \left( \Frac{\phi L}{1-\phi L} - \Frac{\psi^{-1}L}{1-\psi^{-1} L} \right) \\
& = & \Frac{1}{1 - \phi \psi} \left( \Frac{\phi L}{1-\phi L} + \Frac{1}{1-\psi L^{-1} } \right) \\
& = & \Frac{1}{1 - \phi \psi}  \left( \sum_{h=1}^{\infty} \phi^h L^h  + \sum_{h=0}^{\infty} \psi^h L^{-h} \right) \\
& = & \Frac{1}{1 - \phi \psi}  \left( \sum_{h=1}^{\infty} \phi^h L^h  + \sum_{h=0}^{-\infty} \psi^{-h} L^{h} \right),
\end{eqnarray*}

\nin where we observe that the first sum inside the brackets is equal to  $\Frac{1}{1-\phi L} - 1$, and the second sum is equal to $\Frac{1}{1-\psi L^{-1}}$. The moving average coefficients are:

$c_h = \Frac{\phi^h}{1-\phi \psi}$, if $h >0$, \; and \; $c_h = \Frac{\psi^{-h}}{1-\phi \psi}$, if $ h \leq 0,$

\nin since the two formulas coincide for $h=0$ and the convention $0^0=1$ is used (if $\phi$ or $\psi$ is zero).

\bigskip

\nin \textbf{2. Proof of Proposition 2:}

\nin The tail process takes the values either $1/\psi X_{h-1}$, or $\phi X_{h-1}$. Then, we have:
\bigskip

\nin a) if $N+h \leq 0 \iff N \leq - h: X_h = \Frac{1}{1-\phi \psi} \psi^{-N-h}/
\frac{1}{1-\phi \psi} \psi^{-N-h-1} X_{h-1} = \psi^{-1} X_{h-1}$,

\nin b)  if $N+h > 0 \iff N > -h:  X_h = \frac{1}{1-\phi \psi} \phi^{N+h}/
\Frac{1}{1-\phi \psi} \phi^{N+h-1} X_{h-1}= \phi X_{h-1}$,

\nin with $X_0=1$. 

\nin Then, the probability distribution of $N$ is such that :

$P[N=h] = \Frac{(1-\phi \psi)^{-1} \psi^{-h\alpha}}{(1-\phi \psi)^{-1} [\frac{1}{1-\phi^{\alpha}}+ \frac{1}{1-\psi^{\alpha}} -1]}
= \Frac{\psi^{-h \alpha}}{[\frac{1}{1-\phi^{\alpha}}+ \frac{1}{1-\psi^{\alpha}} -1]} $; if $ h \leq 0$,

$P[N=h] = \Frac{(1-\phi \psi)^{-1} \phi^{h\alpha}}{(1-\phi \psi)^{-1} [\frac{1}{1-\phi^{\alpha}}+ \frac{1}{1-\psi^{\alpha}} -1]}
= \Frac{\phi^{h \alpha}}{[\frac{1}{1-\phi^{\alpha}}+ \frac{1}{1-\psi^{\alpha}} -1]} $; if $ h \geq 0$,

\nin with $0^0=1$, by  convention.

\bigskip

\nin \textbf{2a. Proof of Corollary 1}

\nin We proceed in three steps. First, we solve the backward and forward recursive equations satisfied by the sequences $X_h$ starting from the same initial (terminal) condition $X_{-N}$. Next, we determine the value of $X_{-N}$ given the known value $X_0=1$. In the last step, we combine the results.

\nin (i) From Proposition 2 it follows that for $N > -h$ we have :
$$X_h = \phi X_{h-1}$$

\nin Since  $N > -h \iff h > -N \iff h \geq -N+1$, this leads to a recursive formula with the initial condition $X_{-N}$ and the general term

$$X_h = \phi^{h+N} X_{-N}, \; \mbox{for} \; h \geq -N+1$$

\nin (ii) It follows from Proposition 2  that for $N \leq -h$, we have:
$$X_h = \frac{1}{\psi} X_{h-1},$$
\nin which can be written as:
$$X_{h-1} = \psi X_{h}.$$

\nin  Since $N \leq -h \iff -h \geq N \iff h \leq -N$, this leads to a recursive formula with the terminal condition $X_{-N}$ and the general term:

$$X_h = \psi^{-h-N} X_{-N}$$

\nin (iii) These results can be combined, yielding:

$$X_h = ( \phi^{h+N} \id_{N > -h} + \psi^{-h-N} \id_{N \leq -h } ) X_{-N} $$

\nin (iv) The value of $X_{-N}$ is unknown and can be determined from $X_0=1$ by evaluating the above formula for $h=0$. We get:
\begin{eqnarray*}
1=X_0 & = & ( \phi^{N} \id_{N > 0} + \psi^{-N} \id_{N \leq 0} ) X_{-N}\\
& \iff & X_{-N} = \phi^{-N} \id_{N > 0} + \psi^{N} \id_{N \leq 0} 
\end{eqnarray*}

\nin (v) Next, we compute:
\begin{eqnarray*}
X_h & = &   ( \phi^{h+N} \id_{N > -h} + \psi^{-h-N} \id_{N \leq -h } ) ( \phi^{-N} \id_{N > 0} + \psi^{N} \id_{N \leq 0}) \\
& = & \phi^{h}   \id_{N > -h} \id_{N > 0} + \psi^{-h-N} \phi^{-N} \id_{N \leq -h} \id_{N > 0}
+ \phi^{h+N} \psi^{N}  \id_{N > -h} \id_{N \leq 0} + \psi^{-h} \id_{N \leq -h} \id_{N \leq 0} \\
& = & \phi^{h}   \id_{N > Max(-h,0)} +  \psi^{-h-N} \phi^{-N} \id_{0 <N \leq -h} +
\phi^{h+N} \psi^{N}  \id_{-h < N \leq 0} + \psi^{-h} \id_{N \leq Min(-h, 0)}
\end{eqnarray*}
\color{black}
\bigskip

\nin \textbf{3. Tail process representation of AR(2) processes with at least one root inside the unit circle}

\nin The autoregressive process of order 2, i.e. AR(2):

$$y_t = t_1 y_{t-1} + t_2 y_{t-2} + \epsilon_t => y_t (1- t_1L - t_2 L^2)  =  \epsilon_t$$

\nin with i.i.d. errors $\epsilon_t, t=1,2,,..$ and a heavy-tailed error distribution with tail index $\alpha$ can admit a tail process behavior. We distinguish the following cases:

a) $1- t_1L - t_2 L^2 = (1-\lambda_1 L)(1-\lambda_2 L)$, the root reciprocals $\lambda_1$ and $\lambda_2$ are real, distinct, and such that $\lambda_1$ and $\lambda_2$ are real, distinct, and such that $|\lambda_1| <1$ and $|\lambda_2| <1$ (causal process), or $|\lambda_1| >1$ and $|\lambda_2| >1$ (noncausal process). Then, the process can be written as a strictly stationary pure causal or noncausal process.
 
The causal process admits a one-sided MA($\infty$) representation with the coefficients
$c_h = \frac{1}{\lambda_1 + \lambda_2} [(\lambda_2)^{h} + (\lambda_1)^{h}]$, for $h \geq 0$.
The noncausal process admits a one-sided MA($\infty$) representation with coefficients
$c_h = \frac{1}{\lambda_1 -\lambda_2} [(\lambda_2)^{h+1} - (\lambda_1)^{h+1}]$, for $h \leq 0$

\medskip

b) $1- t_1L - t_2 L^2 = (1-\lambda L)^2$, the root reciprocal $\lambda$ is real  and such that either $|\lambda| < 1$ (double causal root), or  $|\lambda| > 1$ (double noncausal root). This case does not satisfy the geometric ergodicity condition and is excluded.

\medskip
d) $1- t_1L - t_2 L^2 = (1-\lambda_1 L)(a-\lambda_2 L)$, the roots $1/\lambda_1$ and $1/\lambda_2$ are imaginary and equal to a pair of complex conjugates of modulus greater than 1. This case is excluded as it results in local explosive oscillations, while a bubble requires local explosive growth.

\bigskip

\nin \textbf{4. Comparison with \cite{fries2022conditional}}
\nin For illustration, let us consider $h=0$ and the MAR(1,1) process with $\alpha$-stable distributed errors using
the approach of \cite{fries2022conditional}. 
It follows from Propositions 1 and 2 that, conditional on a bubble onset at time $t$ and large $y_{t} >y$ we have:

\begin{eqnarray*}
E \left( \frac{u_{t+1}v_{t}}{y_t^2} |y_t>y \right) & = & E \left( \left( \frac{y_{t+1} - \phi y_{t}}{y_t} \right) \left(\frac{y_{t} - \psi y_{t+1}}{y_t}\right) | y_t>y \right)\\
& \approx & E((X_1 - \phi)(1-\psi X_1)) \\
& = & p^+ \left( \psi^{-1} - \phi \right) \left( 1 - \psi \psi^{-1} \right) 
 +  p^{-} \left( \phi - \phi \right) \left( 1 - \psi \phi \right)  \\ 
& = & 0
\end{eqnarray*}

\nin where we replace $\frac{y_{t}}{y_t}$ by $X_0=1$, $\frac{y_{t+1}}{y_t}$ by $X_1$, independent of $t$ and $y$, and the probabilities $p^+$ and $p^{-}$ for $h \leq -1$ and $h \geq 0$, of bubble growth and burst, respectively,
are defined in Proposition 2.

%$p_1=(1-\psi^{-\alpha})\psi^{- \alpha}$ during the growing phase and $p_2 = (1-\phi^{\alpha})\phi^{ \alpha}$ during the decline phase.

Let us consider $h=0$  and the MAR(0,1) process. In this process, component $u_t=1$ and only the component $v_t$ is relevant. Hence, conditional on the bubble onset at time $t$ for large $y_t > y$ we have:

\begin{eqnarray*}
E \left( \frac{u_{t+1}v_{t}}{y_t} |y_t>y \right) & = & E \left(  \left(\frac{y_{t} - \psi y_{t+1}}{y_t}\right) \;| y_t>y \right)\\
& \approx & E (1 - \psi X_1) \\
& = & p^+ \left( 1 - \psi \psi^{-1} \right) \\
& = & 0,
\end{eqnarray*}

\nin where we replace $\Frac{y_{t+1}}{y_t}$ by $X_1$, independent of $t$ and $y$, $\frac{y_{t}}{y_t} = X_0=1$ and the probability  $p^+$ of bubble growth is given in Corollary 2.

\bigskip

\nin \textbf{5. The cumulative probability distribution of $N$}

\nin a) For the MAR(1,1) process:

\nin If $h <0$, we have:

\begin{eqnarray*}
P[N \leq h] & = & \sum_{i \leq h} \Frac{\psi^{-i \alpha}}{[\frac{1}{1-\phi^{\alpha}} +
\frac{1}{1-\psi^{\alpha}} -1]}  \\
& = & \sum_{i \geq -h} \Frac{\psi^{i \alpha}}{[\frac{1}{1-\phi^{\alpha}} +
\frac{1}{1-\psi^{\alpha}} -1]} \\
& = & \Frac{\psi^{-h \alpha}}{1-\psi^{\alpha}} \Frac{1}{[\frac{1}{1-\phi^{\alpha}} +
\frac{1}{1-\psi^{\alpha}} -1]}
\end{eqnarray*}

\nin If $h>0$, then:

$$P[N>h] = \Frac{\phi^{h \alpha}}{1-\phi^{\alpha}} \Frac{1}{[\frac{1}{1-\phi^{\alpha}} +
\frac{1}{1-\psi^{\alpha}} -1]}$$

\medskip

\nin b) For the MAR(0,1) process with $N \leq 0$ and $h\leq 0$, we have:
$$P[N \leq h] = (1-\psi^{\alpha}) \sum_{i \leq h} \psi^{-i \alpha} =(1-\psi^{\alpha}) \sum_{i \geq -h} \psi^{i \alpha} = \psi^{-h \alpha}$$.

\bigskip
\section*{Appendix B}
\label{App:B}

\nin This Appendix describes the conditional covariance of the latent components when the process does not take extreme values. For ease of exposition, consider the MAR(1,1) process defined in Section 2.1.

\nin The latent components $u_t, v_t$  are strictly stationary. When $\epsilon_t$ has finite moments of orders 1 and 2, the latent components have marginal means $E(u_{t}) = E(v_t)= 0$ and their marginal covariance is always known to be zero $E(u_{t+1} v_{t}) = 0$. 

\medskip
Let us consider the conditional moments that exist
regardless of the existence of marginal moments. Since the MAR(1,1) process is Markov of order 2 [see, e.g. \cite{fries2019mixed}, Proposition 3.1], the conditioning set considered is $\underline{y}_t = \{y_t, y_{t-1}\}$. Then $E(u_{t+1}|\underline{y}_{t}) = E [y_{t+1}|\underline{y}_{t}]  - \phi y_{t} $
and
$E(v_{t}|\underline{y}_{t}) = y_{t} - \psi  E [y_{t+1}|\underline{y}_{t}] $.
To interpret the statistic $\xi$, we write the conditional covariance of latent components $E(u_{t+1} v_{t}|\underline{y}_t)$:
$$E(u_{t+1}v_{t} |\underline{y}_t) =  E[(y_{t+1} - \phi y_{t})(y_{t} - \psi  y_{t+1})|\underline{y}_t] = - \phi y_{t}^2  - \psi E[ y_{t+1}^2 |\underline{y}_t]+ (\phi \psi + 1)
y_t E[y_{t+1}|\underline{y}_t].$$

\nin and observe that it depends on the autoregressive coefficients $\phi $ and $\psi$, and the squared values of the process.
\cite{gourieroux2017local} and \cite{fries2019mixed} show that MAR(r,s) processes display conditional heteroskedasticity. More specifically, the conditional volatility of a MAR(1,1) process with Cauchy distributed errors is a quadratic function of the past values of the process, similar to the conditional heteroscedasticity of the Double Autoregressive (DAR) model of \cite{ling2004estimation}
with variance-induced mean reversion. The conditional covariance of the latent components given above depends on the squared past values of the process, and therefore captures the conditional heteroscedasticity of the MAR process, in addition to its serial dependence.

For the MAR (1,1) process with Cauchy-distributed errors and $\psi>0$, the formulas of conditional covariances are easy to compute. We know the equivalence of the conditioning sets $\underline{y}_t = \underline{u}_t$ and from Proposition 5, \cite{gourieroux2017local}, it follows that:
$$E(u_{t+1}|\underline{u}_t) = u_t, \;\; E(u_{t+1}^2|\underline{u}_t) = \frac{1}{\psi} u_t^2 + \frac{\sigma^2}{\psi(1-\psi)}. $$

\nin Since $u_{t+1} = y_{t+1} - \phi y_t$, we can write $y_{t+1} = \phi y_t +u_{t+1}$. Hence,
$$E(y_{t+1}| \underline{y}_t) = E(u_{t+1}| \underline{y}_t) +\phi y_t = u_t + \phi y_t = (y_t - \phi y_{t-1}) + \phi y_t$$

\nin and from Proposition M.1, \cite{fries2019mixed} we get:

$$E[ y_{t+1}^2 |\underline{y}_t] = a y_{t}^2 - 2 b y_{t}y_{t-1} + 
c y_{t-1}^2 + \frac{\sigma^2}{|\psi|(1-|\psi|)}$$

\nin where $a= \phi^2 + 2 \phi \, sign(\psi) + \frac{1}{|\psi|}$, 
$b= \phi^2 sign(\psi) + \frac{\phi}{|\psi|}$ and $c=\frac{\phi^2}{|\psi|}$.
For clarity of exposition, let us assume $\psi>0$. Then, we get:
\begin{eqnarray*}
\lefteqn{E [ (u_{t+1} v_{t})|\underline{y}_{t}] } \\
& = &  - \phi y_{t}^2  - \psi \left(
a y_{t}^2 - 2 b y_{t}y_{t-1} + 
c y_{t-1}^2 +  \frac{\sigma^2}{|\psi|(1-|\psi|)}
\right)
+ (\phi \psi +1) (y_t^2 + (1-L) \phi y_t^2) \\
& =  & y_t^2[(\phi - 1) \phi \psi] + y_t y_{t-1} [2 \phi - \phi(1-\phi \psi)]
- y_{t-1}^2 + \frac{\sigma^2}{(1-|\psi|)}
\end{eqnarray*}

\nin The conditional covariance of the latent components is equal to the constant $\frac{\sigma^2}{(1-|\psi|)}$ when $y_t=y_{t-1}=0$ and it is time-varying otherwise. In particular, for $E[(u_{t+1} v_{t})|\underline{y}_{t}]$ the ratio 

$$E_t\xi_t = \frac{E[(u_{t+1} v_{t})|\underline{y}_{t}]}{y_t^2} \; \mbox{for } 
  y_t \neq 0$$

\nin is time-varying. In general, for $h>1$ we have $E(u_{t+1+h} v_{t+h}|\underline{y}_t)$ where
\begin{eqnarray*}
E(u_{t+1+h}v_{t+h} |\underline{y}_t) & = & E[(y_{t+1+h} - \phi y_{t+h})(y_{t+h} - \psi  y_{t+1+h})|\underline{y}_t] \\
& = & - \phi E [y_{t+h}^2 |\underline{y}_t]  - \psi E[ y_{t+h+1}^2 |\underline{y}_t]+ (\phi \psi + 1)
E[y_{t+h} y_{t+h+1}|\underline{y}_t]
\end{eqnarray*}

\nin The conditional covariance of the latent components is a function of the conditional second moments and the conditional covariance of $y_{t}$ at lag $h$. The first two terms capture the time-varying conditional heteroscedasticity of $y_t$. For positive $\phi, \psi$ these terms become big and negative when the conditioning value is large. 
The third term is the conditional autocovariance at horizon $h$. 
The conditional second moments and conditional autocovariances at lag $h$ can be found from the formula of  $E[y_{t} y_{t+h}| \underline{y}_t]$ given in Proposition M.1, \cite{fries2019mixed} and obtained from the following representation of the process:

$$y_{t+h} = P_h(L) y_{t-1} + Q_h (L^{-1}) u_t,$$

\nin where for $h=0$, we have $y_t = \phi y_{t-1} + u_t$ with $P_0(L) = \phi$ and $Q_0(L^{-1}) = 1$. For $h=1$, we have
$$y_{t+1} = \phi^2 y_{t-1} + \phi u_t + u_{t+1}$$

\nin with $P_1(L) = \phi^2$ and $Q_1(L^{-1}) = \phi+ L^{-1}$. Next, $y_{t+2} = \phi^3 y_{t-1} + u_{t+2} + \phi u_{t+1} + \phi^2 u_t$, with $P_2(L) = \phi^3$ and $Q_2(L^{-1}) = L^{-2} + L^{-1} + \phi^2$, etc.

\newpage
\section{Appendix C}
\label{App:C}

\begin{table}[H]
\centering
\caption{Empirical size of the test for MAR(0,1) estimated by OLS (1000 replications) at 5\% from samples of T=400. Each row corresponds to an error distribution: $t(3)$, $t(4)$, $t(5)$. Columns give the rejection frequency (\emph{Size}) 
for different values of the noncausal autoregressive coefficient $\psi$ (from 0.1 to 0.9). The statistic is evaluated conditional on extreme events, defined as observations exceeding the 97.5th percentile of the simulated series. 
}
\resizebox{\textwidth}{!}{%
\begin{tabular}{lccccccccc}
\toprule
 & \multicolumn{9}{c}{$\psi$} \\
\cmidrule(lr){2-10}
Distribution  & 0.1 & 0.2 & 0.3 & 0.4 & 0.5 & 0.6 & 0.7 & 0.8 & 0.9 \\
\midrule
$t(3)$  & .012 & .012 & .012 & .014 & .015 & .013 & .035 & .053 & .076 \\
$t(4)$  & .050 & .021 & .029 & .027 & .045 & .051 & .069 & .088 & .122 \\
$t(5)$  & .154 & .102 & .056 & .072 & .091 & .085 & .115 & .133 & .138 \\
\bottomrule
\end{tabular}%
}
\end{table}

\begin{table}[H]
\centering
\caption{Empirical power of the test for MAR(0,1) estimated by OLS (1000 replications) at 5\% from samples of T=400. 
Each row corresponds to an error distribution: $t(3)$, $t(4)$, $t(5)$. 
Columns give the rejection frequency (\emph{Power}) for different values of the noncausal autoregressive coefficient $\psi$ (from 0.1 to 0.9). 
The statistic is evaluated at the observation corresponding to the 210th largest value among 400 simulated observations (52.5th percentile). }
\resizebox{\textwidth}{!}{%
\begin{tabular}{lccccccccc}
\toprule
 & \multicolumn{9}{c}{$\psi$} \\
\cmidrule(lr){2-10}
Distribution & 0.1 & 0.2 & 0.3 & 0.4 & 0.5 & 0.6 & 0.7 & 0.8 & 0.9 \\
\midrule
$t(3)$  & .592 & .582 & .555 & .515 & .499 & .531 & .486 & .511 & .496 \\
$t(4)$  & .579 & .589 & .574 & .544 & .541 & .537 & .523 & .502 & .512 \\
$t(5)$  & .561 & .585 & .559 & .525 & .517 & .512 & .500 & .514 & .472 \\
\bottomrule
\end{tabular}%
}
\end{table}

\begin{table}[H]
\centering
\caption{Empirical size and power of the test for MAR(1,1) with Cauchy distributed errors estimated by GCov (1000 replications) at 5\% from samples of T=400.  Columns give the rejection frequency (\emph{Size}) 
for different values of the noncausal autoregressive coefficient $\psi$ (from 0.1 to 0.9). The statistic is evaluated conditional on extreme events, defined as observations exceeding the 97.5th percentile of the simulated series (size), and 210th largest value among 400 simulated observations (52.5th percentile) (power). 
}
\resizebox{\textwidth}{!}{%
\begin{tabular}{ccccccccccccc}
\toprule
\multicolumn{13}{c}{$\psi=0.9$} \\
\hline
$\phi$ & & & &0.1 & 0.2 & 0.3 & 0.4 & 0.5. & 0.6 & 0.7 & 0.8 & 0.9\\ \hline
 size   & & &  &0.08  &  0.05  & 0.04 &  0.08   & 0.04 & 0.06  &  0.04 & 0.01 & 0.01 \\
power  &  & & &0.72  &  0.84  & 0.86 &  0.68   & 0.56 & 0.62  &  0.54 & 0.46 & 0.47 \\
\bottomrule
\end{tabular}%
}
\end{table}

 \begin{table}[htbp]
 \centering
 \caption{Empirical size of the test for MAR(1,1) estimated by GCov (1000 replications) at 5\% from samples of T=400.
 Rows correspond to an error distribution: $t(3)$, $t(4)$, $t(5)$.
 Columns report the rejection frequency (\emph{Size}) for different values of the causal coefficient $\phi$ (0.1--0.9) with $\psi$ fixed. The statistic is evaluated conditional on extreme events, defined as observations exceeding the 97.5th percentile of the simulated series.}
 \label{tab:gcov-mar11-size}
 \resizebox{\textwidth}{!}{%
 \begin{tabular}{l*{9}{c}}
 \toprule
 \multicolumn{10}{c}{$\psi=0.6$} \\
 \hline
 & \multicolumn{9}{c}{$\phi$} \\
 \cmidrule(lr){2-10}
 Distribution & 0.1 & 0.2 & 0.3 & 0.4 & 0.5 & 0.6 & 0.7 & 0.8 & 0.9 \\
 \midrule
 $t(3)$ & .015 & .015 & .012 & .010 & .010 & .012 & .015 & .016 & .012 \\
 $t(4)$ & .006 & .007 & .008 & .006 & .006 & .008 & .011 & .010 & .005 \\
 $t(5)$ & .006 & .004 & .002 & .002 & .002 & .002 & .004 & .004 & .002 \\
 \bottomrule
 \multicolumn{10}{c}{$\psi=0.7$} \\
 \hline
 & \multicolumn{9}{c}{$\phi$} \\
 \cmidrule(lr){2-10}
 Distribution & 0.1 & 0.2 & 0.3 & 0.4 & 0.5 & 0.6 & 0.7 & 0.8 & 0.9 \\
 \midrule
 $t(3)$ & .049 & .064 & .055 & .057 & .049 & .041 & .045 & .050 & .051 \\
 $t(4)$ & .023 & .025 & .033 & .031 & .028 & .023 & .031 & .040 & .036 \\
 $t(5)$ & .016 & .021 & .018 & .016 & .013 & .012 & .012 & .022 & .034 \\
 \bottomrule
 \multicolumn{10}{c}{$\psi=0.8$} \\
 \hline
 & \multicolumn{9}{c}{$\phi$} \\
 \cmidrule(lr){2-10}
 Distribution & 0.1 & 0.2 & 0.3 & 0.4 & 0.5 & 0.6 & 0.7 & 0.8 & 0.9 \\
 \midrule
 $t(3)$ & .056 & .052 & .053 & .052 & .054 & .049 & .046 & .043 & .077 \\
 $t(4)$ & .036 & .033 & .034 & .036 & .040 & .044 & .043 & .038 & .065 \\
 $t(5)$ & .025 & .020 & .028 & .028 & .029 & .027 & .027 & .023 & .052 \\
 \bottomrule
 \multicolumn{10}{c}{$\psi=0.9$} \\
 \hline
 & \multicolumn{9}{c}{$\phi$} \\
 \cmidrule(lr){2-10}
 Distribution & 0.1 & 0.2 & 0.3 & 0.4 & 0.5 & 0.6 & 0.7 & 0.8 & 0.9 \\
 \midrule
 $t(3)$ & .055 & .065 & .072 & .071 & .078 & .080 & .078 & .071 & .064 \\
 $t(4)$ & .045 & .032 & .037 & .043 & .045 & .047 & .042 & .040 & .036 \\
 $t(5)$ & .035 & .045 & .046 & .044 & .051 & .062 & .059 & .052 & .047 \\
 \bottomrule
 \end{tabular}}
 \end{table}

 \begin{table}[htbp]
 \centering
 \caption{ Empirical power of the test for MAR(1,1) estimated by GCov (1000 replications) at 5\% from samples of T=400. 
Each row corresponds to an error distribution: $t(3)$, $t(4)$, $t(5)$. 
Columns give the rejection frequency (\emph{Power}) for different values of the noncausal autoregressive coefficient $\psi$ (from 0.1 to 0.9). 
The statistic is evaluated at the observation corresponding to the 210th largest value among 400 simulated observations (52.5th percentile).} 
%Empirical power of the test for MAR(1,1) estimated by GCov (1000 replications).
% Rows correspond to error distributions ($t(3)$, $t(4)$, $t(5)$).
% Columns report the rejection frequency (\emph{Size}) for different values of the causal coefficient $\phi$ (0.1--0.9) with $\psi$ fixed at 0.9. The statistic is evaluated conditional on extreme events, defined as observations exceeding the 97.5th percentile of the simulated series.
 \label{tab:gcov-mar11-size}
 \resizebox{\textwidth}{!}{%
 \begin{tabular}{l*{9}{c}}
 \toprule
 \multicolumn{10}{c}{$\psi=0.6$} \\
 \hline
 & \multicolumn{9}{c}{$\phi$} \\
 \cmidrule(lr){2-10}
 Distribution & 0.1 & 0.2 & 0.3 & 0.4 & 0.5 & 0.6 & 0.7 & 0.8 & 0.9 \\
 \midrule
  $t(3)$ & .730 & .775 & .783 & .795 & .800 & .790 & .781 & .764 & .759 \\
 $t(4) $& .730 & .785 & .757 & .740 & .756 & .767 & .756 & .738 & .724 \\
 $t(5)$ & .660 & .695 & .707 & .715 & .720 & .713 & .707 & .690 & .681 \\
% $t(3)$ & .113 & .104 & .139 & .156 & .204 & .224 & .331 & .435 & .549 \\
% $t(4) $& .181 & .202 & .233 & .286 & .311 & .386 & .478 & .580 & .623 \\
% $t(5)$ & .217 & .269 & .312 & .381 & .454 & .483 & .556 & .619 & .663 \\
 \bottomrule
 \multicolumn{10}{c}{$\psi=0.7$} \\
 \hline
 & \multicolumn{9}{c}{$\phi$} \\
 \cmidrule(lr){2-10}
 Distribution & 0.1 & 0.2 & 0.3 & 0.4 & 0.5 & 0.6 & 0.7 & 0.8 & 0.9 \\
 \midrule
 $t(3)$ & .740 & .785 & .793 & .797 & .802 & .802 & .791 & .769 & .747 \\
 $t(4)$ & .750 & .750 & .737 & .738 & .740 & .763 & .759 & .744 & .726 \\
 $t(5)$ & .800 & .775 & .763 & .765 & .760 & .762 & .759 & .746 & .723 \\
 \bottomrule
 \multicolumn{10}{c}{$\psi=0.8$} \\
 \hline
 & \multicolumn{9}{c}{$\phi$} \\
 \cmidrule(lr){2-10}
 Distribution & 0.1 & 0.2 & 0.3 & 0.4 & 0.5 & 0.6 & 0.7 & 0.8 & 0.9 \\
 \midrule
 $t(3)$ & .750 & .795 & .793 & .782 & .764 & .763 & .761 & .738 & .714 \\
 $t(4)$ & .800 & .765 & .753 & .752 & .756 & .747 & .739 & .729 & .709 \\
 $t(5)$ & .740 & .775 & .753 & .755 & .752 & .735 & .737 & .731 & .700 \\
 \bottomrule
 \multicolumn{10}{c}{$\psi=0.9$} \\
 \hline
 & \multicolumn{9}{c}{$\phi$} \\
 \cmidrule(lr){2-10}
 Distribution & 0.1 & 0.2 & 0.3 & 0.4 & 0.5 & 0.6 & 0.7 & 0.8 & 0.9 \\
 \midrule
 $t(3)$ & .760 & .725 & .717 & .713 & .706 & .698 & .687 & .676 & .642 \\
 $t(4)$ & .690 & .720 & .683 & .677 & .666 & .670 & .650 & .636 & .611 \\
 $t(5)$ & .720 & .730 & .703 & .688 & .680 & .647 & .641 & .625 & .596 \\
 \bottomrule
 \end{tabular}}
 \end{table}

\newpage

\bibliographystyle{chicago}
\bibliography{references}

@misc{wimmer2016green,
  title={The Green Bubble: Our Future Energy Needs and Why Alternative Energy Is Not the Answer},
  author={Wimmer},
  year={2016},
  publisher={LID Publishing: London, UK}
}

@article{phillips2011explosive,
  title={Explosive behavior in the 1990s Nasdaq: When did exuberance escalate asset values?},
  author={Phillips, Peter CB and Wu, Yangru and Yu, Jun},
  journal={International economic review},
  volume={52},
  number={1},
  pages={201--226},
  year={2011},
  publisher={Wiley Online Library}
}

@incollection{brockwell1987stationary,
  title={Stationary ARMA Processes},
  author={Brockwell, Peter J and Davis, Richard A},
  booktitle={Time Series: Theory and Methods},
  pages={77--111},
  year={1987},
  publisher={Springer}
}

@article{cavaliere2020bootstrapping,
  title={Bootstrapping noncausal autoregressions: with applications to explosive bubble modeling},
  author={Cavaliere, Giuseppe and Nielsen, Heino Bohn and Rahbek, Anders},
  journal={Journal of Business \& Economic Statistics},
  volume={38},
  number={1},
  pages={55--67},
  year={2020},
  publisher={Taylor \& Francis}
}

@article{giancaterini2025regularized,
  title={Regularized Generalized Covariance (RGCov) Estimator},
  author={Giancaterini, Francesco and Hecq, Alain and Jasiak, Joann and Neyazi, Aryan Manafi},
  journal={arXiv preprint arXiv:2504.18678},
  year={2025}
}

@article{phillips2015testing,
  title={Testing for multiple bubbles: Historical episodes of exuberance and collapse in the S\&P 500},
  author={Phillips, Peter CB and Shi, Shuping and Yu, Jun},
  journal={International economic review},
  volume={56},
  number={4},
  pages={1043--1078},
  year={2015},
  publisher={Wiley Online Library}
}

@article{jasiak2023gcov,
  title={GCov-Based Portmanteau Test},
  author={Jasiak, Joann and Neyazi, Aryan Manafi},
  journal={arXiv preprint arXiv:2312.05373},
  year={2023}
}

@article{jasiakhall,
  title={Modelling common bubbles in cryptocurrency prices},
  author={Hall, Mauri K and Jasiak, Joann},
  journal={Economic Modelling},
  volume={139},
  pages={106782},
  year={2024},
  publisher={Elsevier}
}

@article{breid1991maximum,
  title={Maximum likelihood estimation for noncausal autoregressive processes},
  author={Breidt, F Jay and Davis, Richard A and Lh, Keh-Shin and Rosenblatt, Murray},
  journal={Journal of Multivariate Analysis},
  volume={36},
  number={2},
  pages={175--198},
  year={1991},
  publisher={Elsevier}
}

@article{lanne2011noncausal,
  title={Noncausal autoregressions for economic time series},
  author={Lanne, Markku and Saikkonen, Pentti},
  journal={Journal of Time Series Econometrics},
  volume={3},
  number={3},
  year={2011},
  publisher={De Gruyter}
}

@article{hecq2016identification,
  title={Identification of mixed causal-noncausal models in finite samples},
  author={Hecq, Alain and Lieb, Lenard and Telg, Sean},
  journal={Annals of Economics and Statistics/Annales d'{\'E}conomie et de Statistique},
  number={123/124},
  pages={307--331},
  year={2016},
  publisher={JSTOR}
}

@article{cubadda2023optimization,
  title={Optimization of the Generalized Covariance Estimator in Noncausal Processes},
  author={Cubadda, Gianluca and Giancaterini, Francesco and Hecq, Alain and Jasiak, Joann},
  journal={arXiv preprint arXiv:2306.14653},
  year={2023}
}

@article{gourieroux2017noncausal,
  title={Noncausal vector autoregressive process: Representation, identification and semi-parametric estimation},
  author={Gourieroux, Christian and Jasiak, Joann},
  journal={Journal of Econometrics},
  volume={200},
  number={1},
  pages={118--134},
  year={2017},
  publisher={Elsevier}
}

@article{gourieroux2023generalized,
  title={Generalized covariance estimator},
  author={Gourieroux, Christian and Jasiak, Joann},
  journal={Journal of Business \& Economic Statistics},
  volume={41},
  number={4},
  pages={1315--1327},
  year={2023},
  publisher={Taylor \& Francis}
}

@article{lanne2013noncausal,
  title={Noncausal vector autoregression},
  author={Lanne, Markku and Saikkonen, Pentti},
  journal={Econometric Theory},
  volume={29},
  number={3},
  pages={447--481},
  year={2013},
  publisher={Cambridge University Press}
}

@article{gourieroux2017local,
  title={Local explosion modelling by non-causal process},
  author={Gourieroux, Christian and Zakoian, Jean-Michel},
  journal={Journal of the Royal Statistical Society Series B: Statistical Methodology},
  volume={79},
  number={3},
  pages={737--756},
  year={2017},
  publisher={Oxford University Press}
}

@article{fries2019mixed,
  title={Mixed causal-noncausal ar processes and the modelling of explosive bubbles},
  author={Fries, Sebastien and Zakoian, Jean-Michel},
  journal={Econometric Theory},
  volume={35},
  number={6},
  pages={1234--1270},
  year={2019},
  publisher={Cambridge University Press}
}

@article{hecq2021forecasting,
  title={Forecasting bubbles with mixed causal-noncausal autoregressive models},
  author={Hecq, Alain and Voisin, Elisa},
  journal={Econometrics and Statistics},
  volume={20},
  pages={29--45},
  year={2021},
  publisher={Elsevier}
}

@article{giancaterini2022climate,
  title={Is climate change time-reversible?},
  author={Giancaterini, Francesco and Hecq, Alain and Morana, Claudio},
  journal={Econometrics},
  volume={10},
  number={4},
  pages={36},
  year={2022},
  publisher={MDPI}
}

@article{davis2020noncausal,
  title={Noncausal vector AR processes with application to economic time series},
  author={Davis, Richard A and Song, Li},
  journal={Journal of Econometrics},
  volume={216},
  number={1},
  pages={246--267},
  year={2020},
  publisher={Elsevier}
}

@article{chan2006note,
  title={A note on time-reversibility of multivariate linear processes},
  author={Chan, Kung-Sik and Ho, Lop-Hing and Tong, Howell},
  journal={Biometrika},
  volume={93},
  number={1},
  pages={221--227},
  year={2006},
  publisher={Oxford University Press}
}

@article{lanne2012optimal,
  title={Optimal forecasting of noncausal autoregressive time series},
  author={Lanne, Markku and Luoto, Jani and Saikkonen, Pentti},
  journal={International Journal of Forecasting},
  volume={28},
  number={3},
  pages={623--631},
  year={2012},
  publisher={Elsevier}
}

@article{gourieroux2015uniqueness,
  title={\textcolor{blue}{On Uniqueness of Moving Average Representations of Heavy-tailed Stationary Processes}},
  author={Gourieroux, Christian and Zakoian, Jean-Michel},
  journal={\textcolor{blue}{Journal of Time Series Analysis}},
  volume={\textcolor{blue}{36}},
  number={\textcolor{blue}{6}},
  pages={\textcolor{blue}{876--887}},
  year={2015},
  publisher={\textcolor{blue}{Wiley Online Library}}
}

@article{gourieroux2016filtering,
  title={Filtering, prediction and simulation methods for noncausal processes},
  author={Gourieroux, Christian and Jasiak, Joann},
  journal={Journal of Time Series Analysis},
  volume={37},
  number={3},
  pages={405--430},
  year={2016},
  publisher={Wiley Online Library}
}

@article{giorgis2024salvation,
  title={‘Salvation and profit’: deconstructing the clean-tech bubble},
  author={Giorgis, Vincent and Huber, Tobias A and Sornette, Didier},
  journal={Technology Analysis \& Strategic Management},
  volume={36},
  number={4},
  pages={827--839},
  year={2024},
  publisher={Taylor \& Francis}
}

@article{fries2022conditional,
  title={Conditional moments of noncausal alpha-stable processes and the prediction of bubble crash odds},
  author={Fries, S{\'e}bastien},
  journal={Journal of Business \& Economic Statistics},
  volume={40},
  number={4},
  pages={1596--1616},
  year={2022},
  publisher={Taylor \& Francis}
}

@book{hull2016options,
  title={Options, futures, and other derivatives},
  author={Hull, John C and Basu, Sankarshan},
  year={2016},
  publisher={Pearson Education India}
}

@article{sadorsky2012correlations,
  title={Correlations and volatility spillovers between oil prices and the stock prices of clean energy and technology companies},
  author={Sadorsky, Perry},
  journal={Energy economics},
  volume={34},
  number={1},
  pages={248--255},
  year={2012},
  publisher={Elsevier}
}

@article{sadorsky2012modeling,
  title={Modeling renewable energy company risk},
  author={Sadorsky, Perry},
  journal={Energy Policy},
  volume={40},
  pages={39--48},
  year={2012},
  publisher={Elsevier}
}

@article{henriques2008oil,
  title={Oil prices and the stock prices of alternative energy companies},
  author={Henriques, Irene and Sadorsky, Perry},
  journal={Energy Economics},
  volume={30},
  number={3},
  pages={998--1010},
  year={2008},
  publisher={Elsevier}
}

@article{kumar2012stock,
  title={Stock prices of clean energy firms, oil and carbon markets: A vector autoregressive analysis},
  author={Kumar, Surender and Managi, Shunsuke and Matsuda, Akimi},
  journal={Energy Economics},
  volume={34},
  number={1},
  pages={215--226},
  year={2012},
  publisher={Elsevier}
}

@article{managi2013does,
  title={Does the price of oil interact with clean energy prices in the stock market?},
  author={Managi, Shunsuke and Okimoto, Tatsuyoshi},
  journal={Japan and the world economy},
  volume={27},
  pages={1--9},
  year={2013},
  publisher={Elsevier}
}

@article{khalifa2022accelerating,
  title={Accelerating the transition to a circular economy for net-zero emissions by 2050: a systematic review},
  author={Khalifa, Ahmed A and Ibrahim, Abdul-Jalil and Amhamed, Abdulkarem I and El-Naas, Muftah H},
  journal={Sustainability},
  volume={14},
  number={18},
  pages={11656},
  year={2022},
  publisher={MDPI}
}

@article{gourieroux2015pricing,
  title={Pricing with finite dimensional dependence},
  author={Gourieroux, Christian and Monfort, Alain},
  journal={Journal of Econometrics},
  volume={187},
  number={2},
  pages={408--417},
  year={2015},
  publisher={Elsevier}
}

@article{mohammed2023all,
  title={Do all renewable energy stocks react to the war in Ukraine? Russo-Ukrainian conflict perspective},
  author={Mohammed, Kamel Si and Usman, Muhammad and Ahmad, Paiman and Bulgamaa, Urangoo},
  journal={Environmental Science and Pollution Research},
  volume={30},
  number={13},
  pages={36782--36793},
  year={2023},
  publisher={Springer}
}

@article{hencic2015noncausal,
  title={Noncausal autoregressive model in application to bitcoin/USD exchange rates},
  author={Hencic, Andrew and Gouri{\'e}roux, Christian},
  journal={Econometrics of risk},
  pages={17--40},
  year={2015},
  publisher={Springer}
}

@article{lof2017noncausality,
  title={Noncausality and the commodity currency hypothesis},
  author={Lof, Matthijs and Nyberg, Henri},
  journal={Energy Economics},
  volume={65},
  pages={424--433},
  year={2017},
  publisher={Elsevier}
}

@article{cubadda2011testing,
  title={Testing for common autocorrelation in data-rich environments},
  author={Cubadda, Gianluca and Hecq, Alain},
  journal={Journal of Forecasting},
  volume={30},
  number={3},
  pages={325--335},
  year={2011},
  publisher={Wiley Online Library}
}

@article{ling2004estimation,
  title={Estimation and testing stationarity for double-autoregressive models},
  author={Ling, Shiqing},
  journal={Journal of the Royal Statistical Society Series B: Statistical Methodology},
  volume={66},
  number={1},
  pages={63--78},
  year={2004},
  publisher={Oxford University Press}
}

@book{kulik2020heavy,
  title={Heavy-tailed time series},
  author={Kulik, Rafal and Soulier, Philippe},
  year={2020},
  publisher={Springer}
}

@article{basrak2009regularly,
  title={Regularly varying multivariate time series},
  author={Basrak, Bojan and Segers, Johan},
  journal={Stochastic processes and their applications},
  volume={119},
  number={4},
  pages={1055--1080},
  year={2009},
  publisher={Elsevier}
}

@incollection{hecq2023predicting,
  title={Predicting crashes in oil prices during the COVID-19 pandemic with mixed causal-noncausal models},
  author={Hecq, Alain and Voisin, Elisa},
  booktitle={Essays in honor of Joon Y. Park: Econometric methodology in empirical applications},
  volume={45},
  pages={209--233},
  year={2023},
  publisher={Emerald Publishing Limited}
}

@article{Blasq,
title={A Novel Test for the Presence of Local Explosive Dynamics},
author={Blasques, F and Koopman, S. J and  Mingoli, G and S Telg},
journal={Journal of Time Series Analysis},
volume={46},
pages={966-980},
year={2025}
}

@article{de2025forecasting,
  title={Forecasting Extreme Trajectories Using Seminorm Representations},
  author={de Truchis, Gilles and Fries, S{\'e}bastien and Thomas, Arthur},
journal={Document de Recherche du Laboratoire d’Économie d’Orléans},
  year={2025}
}

@article{davis2018inference,
  title={\textcolor{blue}{Inference on the tail process with application to financial time series modeling}},
  author={Davis, Richard A and Drees, Holger and Segers, Johan and Warcho{\l}, Micha{\l}},
  journal={\textcolor{blue}{Journal of Econometrics}},
  volume={\textcolor{blue}{205}},
  number={\textcolor{blue}{2}},
  pages={\textcolor{blue}{508--525}},
  year={\textcolor{blue}{2018}},
  publisher={\textcolor{blue}{Elsevier}}
}

@techreport{Blasques2023Observation,
author = {Francisco Blasques and Siem Jan Koopman and Gabriele Mingoli},
copyright = {\textcolor{blue}{https://www.econstor.eu/dspace/Nutzungsbedingungen}},
language = {eng},
number = {TI 2023-065/III},
publisher = {\textcolor{blue}{Tinbergen Institute}},
title = {\textcolor{blue}{Observation-Driven filters for Time-Series with Stochastic Trends and Mixed Causal Non-Causal Dynamics}},
type = {\textcolor{blue}{Tinbergen Institute Discussion Paper}},
url = {\textcolor{blue}{https://hdl.handle.net/10419/282878}},
year = {\textcolor{blue}{2023}}
}

@article{virani2019sequential,
  title={\textcolor{blue}{Sequential hypothesis tests for streaming data via symbolic time-series analysis}},
  author={Virani, Nurali and Jha, Devesh K and Ray, Asok and Phoha, Shashi},
  journal={\textcolor{blue}{Engineering Applications of Artificial Intelligence}},
  volume={\textcolor{blue}{81}},
  pages={\textcolor{blue}{234--246}},
  year={\textcolor{blue}{2019}},
  publisher={\textcolor{blue}{Elsevier}}
}

@article{gourieroux2026nonlinear,
  title={\textcolor{blue}{Nonlinear Fore (Back) Casting and Innovation Filtering for Causal--Noncausal VAR Models}},
  author={Gourieroux, Christian and Jasiak, Joann},
  journal={\textcolor{blue}{Journal of Financial Econometrics}},
  volume={\textcolor{blue}{24}},
  number={\textcolor{blue}{2}},
  pages={\textcolor{blue}{nbag005}},
  year={\textcolor{blue}{2026}},
  publisher={\textcolor{blue}{Oxford University Press}}
}

@article{drees2020peak,
  title={\textcolor{blue}{Peak-over-threshold estimators for spectral tail processes: random vs deterministic thresholds}},
  author={Drees, Holger and Kne{\v{z}}evi{\'c}, Miran},
  journal={\textcolor{blue}{Extremes}},
  volume={\textcolor{blue}{23}},
  number={\textcolor{blue}{3}},
  pages={\textcolor{blue}{465--491}},
  year={\textcolor{blue}{2020}},
  publisher={\textcolor{blue}{Springer}}
}

\end{document}